\documentclass[apj,twocolumn]{openjournal}
\usepackage{amsmath}
\usepackage{booktabs}
\usepackage{multirow}
\usepackage{color}
\usepackage{soul}

\usepackage[dvipsnames]{xcolor} 

\newcommand{\ztfbd}{ZTF J2020+5033 }
\newcommand{\ztfb}{ZTF J2020+5033}

\usepackage[breaklinks,colorlinks,citecolor=blue,urlcolor=blue]{hyperref}

\usepackage{orcidlink}

\renewcommand{\arraystretch}{1.2}  
\usepackage{listings}
\usepackage{color}
\definecolor{dkgreen}{rgb}{0,0.6,0}
\definecolor{gray}{rgb}{0.5,0.5,0.5}
\definecolor{mauve}{rgb}{0.58,0,0.82}
\definecolor{golden}{rgb}{0.86,0.65,0.01}
\begin{document}


\title{A transiting brown dwarf in a 2 hour orbit }

\author{\vspace{-1.3cm}Kareem El-Badry\,\orcidlink{0000-0002-6871-1752}$^{1}$}
\author{Kevin B. Burdge\,\orcidlink{0000-0002-7226-836X}$^{2}$}
\author{Jan van Roestel\,\orcidlink{0000-0002-2626-2872}$^{3}$}
\author{Antonio C. Rodriguez\,\orcidlink{0000-0003-4189-9668}$^{1}$}

\affiliation{$^1$Department of Astronomy, California Institute of Technology, 1200 E. California Blvd., Pasadena, CA 91125, USA}
\affiliation{$^2$MIT Kavli Institute for Astrophysics and Space Research, 77 Massachusetts Ave., Cambridge, MA 02139, USA}
\affiliation{$^3$Anton Pannekoek Institute for Astronomy, University of Amsterdam, 1090 GE Amsterdam, The Netherlands}
\email{Corresponding author: kelbadry@caltech.edu}

\begin{abstract}
We report the discovery of \ztfb, a high-mass brown dwarf (BD) transiting a low-mass star with an orbital period of 1.90 hours. Phase-resolved spectroscopy, optical and infrared light curves, and precise astrometry from {\it Gaia} allow us to constrain the masses, radii, and temperatures of both components with few-percent precision. We infer a BD mass of $M_{\rm BD} = 80.1\pm 1.6\,M_{\rm J}$, almost exactly at the stellar/substellar boundary, and a moderately inflated radius, $R_{\rm BD} = 1.05\pm 0.02\,R_{\rm J}$. 
The transiting object's temperature, $T_{\rm eff}\approx 1700\,\rm K$, is well-constrained by the depth of the infrared secondary eclipse and strongly suggests it is a BD. The system's high tangential velocity ($v_\perp = 98\,\rm km\,s^{-1}$) and thick disk-like Galactic orbit imply the binary is old; its close distance ($d\approx 140$\,pc) suggests that BDs in short-period orbits are relatively common.  \ztfbd is the shortest-period known transiting BD by more than a factor of 7. Today, the entire binary would comfortably fit inside the Sun. However, both components must have been considerably larger in youth, implying that the orbit has shrunk by at least a factor of $\sim 5$ since formation. The simplest explanation is that magnetic braking continues to operate efficiently in at least some low-mass stars and BDs.

\end{abstract}

\maketitle

\section{Introduction}
\label{sec:intro}
Brown dwarfs (BDs) are degenerate substellar objects with masses ranging roughly from 13 to 80 $M_{\rm J}$, between giant planets and stars, and radii similar to Jupiter. The temperatures in BD cores are not high enough to sustain long-term hydrogen fusion, so BDs cool as they age, slowly radiating away energy, contracting, and fading. 

A large majority of the $\sim 5000$ known BDs are isolated and were discovered via their thermal emission in infrared surveys \citep[e.g.][]{Gelino2009, Mace2014, Kirkpatrick2021, Aganze2022}. BDs in binaries are surprisingly rare: only about 1\% of solar-type and lower-mass stars have BD companions within a few au, making BD companions significantly less common than either higher-mass stellar companions or lower-mass planets \citep{Marcy2000, Grether2006, Raghavan2010, Triaud2017}. Several theoretical explanations for this ``brown dwarf desert'' have been proposed, with most invoking different formation mechanisms for close-in giant planets, BDs, and low-mass stars \citep[e.g.][]{Stamatellos2009, Ma2014}.

BDs in binaries are sought after both as probes of the star formation process and as laboratories in which to measure BD masses and radii. 
About 50 transiting BDs have been discovered in the last decade \citep[see][for a recent summary]{Carmichael2023}. Most of these systems were initially discovered as transiting planet candidates by surveys including WASP \citep{Pollacco2006}, {\it CoRoT} \citep{Baglin2006}, {\it Kepler} \citep{Borucki2010}, and {\it TESS} \citep{Ricker2015}, and were identified as BDs after radial velocity (RV) follow-up. 
Radius measurements of transiting BDs have revealed significant scatter in the distribution of observed radii at fixed mass, even among BDs with similar ages and compositions \citep[e.g.][]{Bouchy2010, Hodzic2018, Carmichael2020,  Casewell2020, Casewell2020b, Acton2021}. On average, observed transiting BDs are 5-10\% larger than predicted by models. The degree of this ``radius inflation'' is only weakly correlated with irradiation from the BDs' stellar companions \citep{Casewell2020, Sainsbury-Martinez2021}. It is  reminiscent of the radius inflation observed in many M dwarfs \citep{Lopez-Morales2005, Bayless2006, Irwin2009, Cruz2018, Kesseli2018, Parsons2018, Jaehnig2019, Jackson2019}, which has been attributed to a variety of processes including magnetic inhibition of convection, flux-blocking by starspots, and tidal heating.

Most of the known transiting BDs have orbital periods between 1 and 10 days and transit solar-type main sequence stars. The shortest-period BD + main sequence binary discovered so far has an orbital period of 0.56 days \citep{Parviainen2020}. It is uncertain whether BDs are expected to exist in significantly shorter orbits. On the one hand, BDs can have high densities and can -- if paired with another BD or low-mass star -- fit into orbits with periods as short as $\approx 40$ minutes \citep[e.g.][]{Rappaport2021}. On the other, evolutionary models struggle to explain the formation of short-period BD and M dwarf binaries: most models assume that magnetic braking --  which is thought to be the dominant mechanism through which low-mass binaries can lose angular momentum after their formation -- is inefficient below the fully-convective boundary \citep[e.g.][]{Rappaport1983, Stepien2006, Reiners2008, Schreiber2010, Garraffo2015}, and thus that low-mass binary orbits should not shrink significantly after formation. Since BDs and low-mass stars have $\gtrsim 1\,R_\odot$ radii at young ages \citep[$\lesssim 1$ Myr, e.g.][]{Chabrier1997, Phillips2020}, they cannot reach $P_{\rm orb} \lesssim 1$\,day without some angular momentum loss mechanism.

Magnetic braking is a process through which stellar winds carry away angular momentum when magnetic fields cause them to co-rotate with the stellar surface \citep[e.g.][]{Schatzman1962}. In close binaries, tides keep the component stars rotating on the orbital period, so the angular momentum is extracted from the orbit, causing it to shrink. For reasons primarily related to the observed mass and period distributions of close white dwarf + main sequence binaries \citep[e.g.][]{Rappaport1983, Schreiber2010}, it is often assumed that magnetic braking becomes inefficient below the fully convective boundary, corresponding roughly to $M \lesssim 0.35\,M_{\odot}$.

It is tempting to attribute the lack of observed short-period BDs and low-mass stars to inefficient magnetic braking in fully-convective stars \citep[e.g.][]{Stepien2006}.  It is worth remembering, however, that observational selection effects disfavor short-period binaries in magnitude-limited samples. The lack of detected short-period transiting BDs thus may simply reflect the fact that any main-sequence star or BD that can fit inside short-period orbits must necessarily be faint \citep[e.g.][]{El-Badry2022}.

This paper presents the discovery of a BD in a binary with an orbital period much shorter than any such system discovered to date. The discovery was enabled by the Zwicky Transient Facility \citep[ZTF;][]{Bellm2019}, which provides high-quality light curves for stars significantly fainter than other surveys that have been used to detect transiting BDs thus far. The remainder of the paper is organized as follows. Section~\ref{sec:disc} describes the object's discovery and our follow-up photometric and spectroscopic observations. Section~\ref{sec:jointmodel} presents our modeling of the system and a joint fit of its multi-band light curves, RVs, astrometry, and broadband SED. In Section~\ref{sec:discussion}, we compare the system to other known binaries containing BDs and discuss its formation history and future evolution. We summarize our results in Section~\ref{sec:conclusion}.

\section{Discovery and follow-up observations}
\label{sec:disc}
We discovered \ztfbd in the course of a search for low-mass eclipsing binaries described by \citet{El-Badry2022}. That work applied the box least squares algorithm \citep{Kovacs2002} to ZTF light curves of main-sequence stars within $\sim 500$ pc of the Sun. The search identified 469 eclipsing binaries with extinction-corrected $G-$band absolute magnitudes $M_{G,0} > 10$, a limit that was chosen to select binaries in which both components are fully convective. 

Visual inspection of the 469 light curves revealed \ztfbd to be unique in that its optical light curves show a strong primary eclipse but no secondary eclipse. This implied that one component was much cooler than the other. Given that the unresolved source falls near the bottom of the main sequence and the primary is already quite cool (Figure~\ref{fig:summary}), this suggested that the companion might be a BD, prompting our follow-up observations. 

Basic observables of the system are shown in Figure~\ref{fig:summary}, and its parameters are summarized in Table~\ref{tab:system}. The unresolved source has apparent magnitude $G=18.7$, absolute magnitude $M_G=12.92$, and colors $G - G_{\rm RP}=1.30$ and $G_{\rm BP}-G_{\rm RP}=3.37$. The {\it Gaia} parallax implies a distance of $142.7\pm 3$\,pc and places the source on the lower main sequence, well below the fully convective boundary. Assuming only one component contributes significantly to the optical photometry, the observed color and absolute magnitude imply a spectral type between M5 and M6 \citep[e.g.][]{Smart2021}. Unlike most of the eclipsing binaries in the \citet{El-Badry2022} sample, the source is not above the main sequence in the color-magnitude diagram. The parameters that we infer from light curve modeling (Section~\ref{sec:jointmodel}) imply that the M dwarf dominates in the optical, but the BD contributes $\approx 10\%$ of the light in the infrared; their separate and combined spectral energy distributions are shown in the right panel of Figure~\ref{fig:summary}.

\subsection{Light curves}
\label{sec:light_curves}

\subsubsection{ZTF}
We analyzed the source's ZTF DR17 light curves, which contain 881 clean epochs (i.e., \texttt{catflags} = 0) in the $r-$ band and 141 in the $g-$ band. The $g-$band data is noisy since the source is quite red, so we focus our analysis on the $r-$band data. The light curve spans 4.7 years ($\approx 21600$ orbits), with 30s exposures and a median uncertainty of 0.07 mag.

\begin{figure*}
    \centering
    \includegraphics[width=0.9\textwidth]{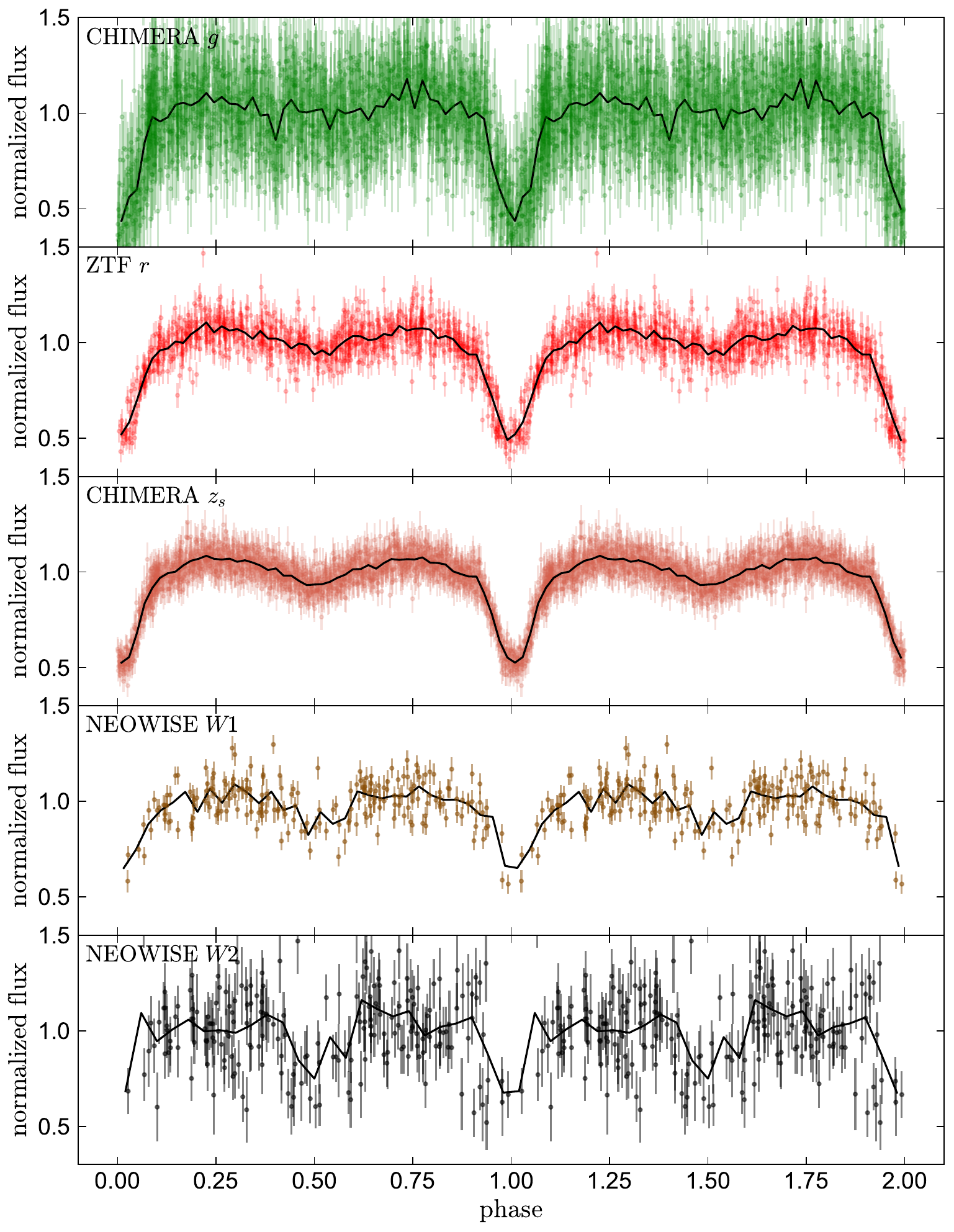}
    \caption{Phased light curves of \ztfbd in five different bandpasses, with wavelength increasing from top to bottom. In the optical, there is a deep primary eclipse but no secondary eclipse, indicating that the secondary is much cooler than the primary. Secondary eclipses become evident in the infrared, where the cool secondary contributes a significant fraction of the total light. This constrains the secondary temperature (Figure~\ref{fig:fitting_results}).}
    \label{fig:lcs}
\end{figure*}

The orbital period, $P_{\rm orb} \approx 1.90$ hr, was first estimated from the box-least-squares periodogram and then refined by fitting a light curve model (section~\ref{sec:ephemeris}). In addition to an obvious primary eclipse, the phased $r-$band light curve shows sinusoidal modulation on half the orbital period (Figure~\ref{fig:lcs}). We attribute this to tidal deformation of the M dwarf, allowing us to rule out a scenario where the period is 3.8 hours and the primary and secondary eclipses have similar depths. 

\begin{figure*}
    \centering
    \includegraphics[width=\textwidth]{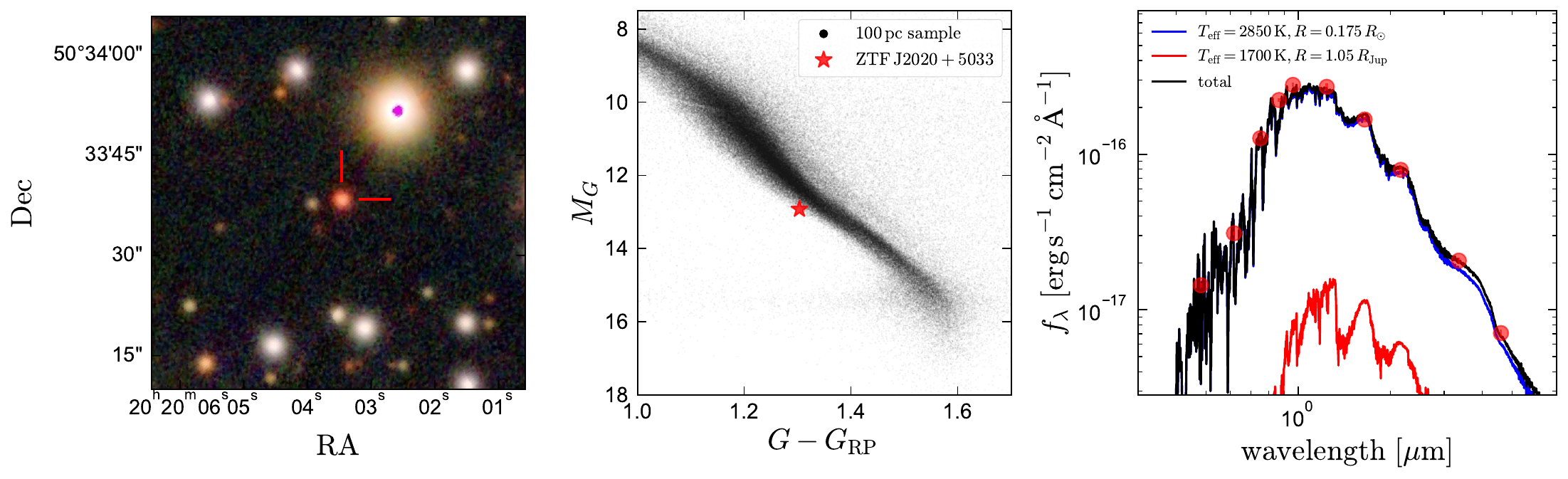}
    \caption{Left: 1-arcminute wide Pan-STARRS $i/r/g$ postage stamp centered on \ztfb. Middle: the source's position on the {\it Gaia} color-magnitude diagram, compared to the 100 pc sample. The source is slightly blueward of the main sequence, suggesting low metallicity and little contribution from a second star in the optical. Right: optical and infrared spectral energy distribution (SED), with best-fit models (Section~\ref{sec:jointmodel}) overplotted. The SED is dominated by an M dwarf with $T_{\rm eff} \approx 2850\, \rm K$ and $R\approx 0.175\,R_{\odot}$. In the infrared, the cooler brown dwarf contributes $\approx 10\%$ of the light.  }
    \label{fig:summary}
\end{figure*}

\subsubsection{CHIMERA}
\label{sec:chimera}
We observed \ztfbd for a full orbit on 2022 August 25 with the Caltech HIgh-speed Multi-color camERA \citep[CHIMERA;][]{chimera}. We observed in the Sloan $g$-and $z_s$-bands simultaneously, using 5 second exposures with $\approx 0.5$ second gaps between frames. Both light curves are shown in Figure~\ref{fig:lcs}. The signal-to-noise ratio (SNR) of the $z_s$-band data is much higher than that of the $g-$band data. The primary eclipse is slightly deeper in the $g-$band. Ellipsoidal variation and hints of a secondary eclipse are evident in the $z_s-$band data.

\subsubsection{WISE}
\label{sec:wise}
\ztfbd has been observed regularly by the {\it WISE} satellite \citep{Wright2010, Mainzer2014}. We retrieved the NEOWISE light curve from IRSA/IPAC, restricting our analysis to data with \texttt{qual\_frame > 0}. This yielded 208 usable measurements in the $W1$ ($3.4\,\mu$m) and $W2$ ($4.6\,\mu$m) bands. The light curve spans more than 8 years, with groups of 10-15 observations every $\sim 6$ months and exposures within each group typically separated by a few hours, resulting in fairly uniform phase coverage. 

The {\it WISE} light curves are shown in the bottom two panels of Figure~\ref{fig:lcs}. Unlike in the optical data, there is clear evidence of a secondary eclipse at phase 0.5. This suggests that the secondary contributes a significant fraction of the total light in the infrared, but not in the optical, as expected if it has a comparable radius to the primary but a cooler temperature.

\subsubsection{Photometric ephemeris}
\label{sec:ephemeris}
We fit the ZTF $r-$band light curve to determine a precise ephemeris as follows. We construct a light curve model for the $r-$band light curve using  \texttt{ellc} \citep{Maxted2016} and the parameters inferred in Section~\ref{sec:jointmodel}. Treating $P_{\rm orb}$ and $t_0$ as free parameters, we predict the phase of each measurement and the corresponding predicted flux at that phase. We sample from the posteriors of $P_{\rm orb}$ and $t_0$ using \texttt{emcee} \citep{Foreman-Mackey2013}, employing flat priors and a likelihood that compares the predicted and measured normalized fluxes, assuming Gaussian uncertainties. We choose the orbital cycle on which $t_0$ is constrained to coincide with the ESI spectroscopic observations and adopt a convention where $t_0$ corresponds to the primary eclipse (i.e., BD in front of M dwarf).

The resulting constraints on $P_{\rm orb}$ and $t_0$ are reported in Table~\ref{tab:system}. The orbital period, $P_{\rm orb}=6850.1653\pm 0.0014$ seconds, or $1.90282368 \pm 0.00000038$ hours, is constrained to 2 parts in $10^7$. This in principle allows the phase to be predicted with $1\%$ accuracy within $\pm 10$ years of $t_0$. However, we caution that many close binaries undergo orbital period modulations due to magnetic activity in the component stars \citep{Applegate1992, Lanza1998}, and the CHIMERA light curve suggests that such variations indeed occur in \ztfbd 
(Section~\ref{sec:ttvs}). The predictive power of the ephemeris is thus likely somewhat lower than the formal uncertainties suggest.

\subsubsection{Evidence of transit timing variations}
\label{sec:ttvs}
When we phased the CHIMERA data to the ephemeris inferred from the ZTF light curve, we found that the primary eclipse occurred $45\pm 3$ seconds later than expected. This delay, while small, is quite significant, and the light curve fit is visibly poor if it is not accounted for. The simplest explanation is that the binary undergoes orbital period modulations resulting from magnetic activity-driven fluctuations in the components' quadrupole moments, as is common in close binaries \citep[e.g.][]{Applegate1992, Lanza1998, Watson2010}. 

It is also possible that the delay is a result of changes in light travel time due to the orbit of the binary around an unseen tertiary. In this case, a 45 second delay would correspond to a displacement of the binary by 0.09 au along the line of sight. There are no nearby resolved {\it Gaia} sources whose parallaxes and proper motions are consistent with being bound to \ztfbd \citep{El-Badry2021}. Given the faintness of the source and lack of spectral contributions from another luminous source (Section~\ref{sec:esi}), the only plausible companions would be BDs or circumbinary planets. Such companions cannot be excluded, but given the frequent occurrence of timing variations in close binaries due to magnetic activity, the observed delay does not in itself provide strong evidence for a third body.

\subsection{Phase-resolved spectroscopy}
\label{sec:esi}
We observed \ztfbd for $\sim 2$ hours on 2022 June 3 with the Echellette Spectrograph and Imager \citep[ESI;][]{Sheinis2002} on the 10\,m Keck-II telescope on Maunakea. We used the $0.75''$ slit and $2\times 1$ binning, yielding a resolution $R\approx 5500$, with wavelength coverage of 3900--10,000\,\AA. We reduced the data using the MAuna Kea Echelle Extraction (MAKEE) pipeline, which performs bias subtraction, flat fielding, wavelength calibration, and sky subtraction. We obtained 19 exposures, each with exposure time 300 seconds, and a 54 second gap between exposures for read-out (Table~\ref{tab:obslog}). 

We flux-calibrated and merged spectra from individual orders using observations of a flux standard taken the same night. Small shifts in the wavelength solution were corrected by fitting the telluric absorption lines with a HITRAN model \citep{Rothman2009, Gullikson2014}; the inferred shifts were always within $\pm 2\,\rm km\,s^{-1}$ of 0. To correct for slit losses and atmospheric dispersion, we fit a first-order polynomial correction to the flux calibration by requiring that the orbit-averaged coadded spectrum is consistent with the Pan-STARRS photometry.

Cutouts of the phased spectra are shown in the top panels of Figure~\ref{fig:river}. All the photospheric lines vary sinusoidally, as expected for a single-lined binary. There is no evidence of a second luminous component moving in anti-phase. H$\alpha$ and H$\beta$  are both observed in emission; this is most likely a result of chromospheric activity in the rapidly-rotating M dwarf \citep[e.g.][]{Newton2017}.

\subsubsection{Radial velocities}
We measured RVs for the M dwarf by cross-correlating the normalized, tellurics-corrected spectra with a \texttt{BT-Settl} model spectrum \citep{Allard2011}. We used a solar-metallicity model with $T_{\rm eff}=2900\,\rm K$ and $\log \left[\left(g/{\rm cm\,s^{-1}}\right)\right]= 5$, consistent with our fitting of the spectrum and SED (Sections~\ref{sec:sptype} and \ref{sec:jointmodel}). 

\begin{figure*}[!ht]
    \centering
    \includegraphics[width=\textwidth]{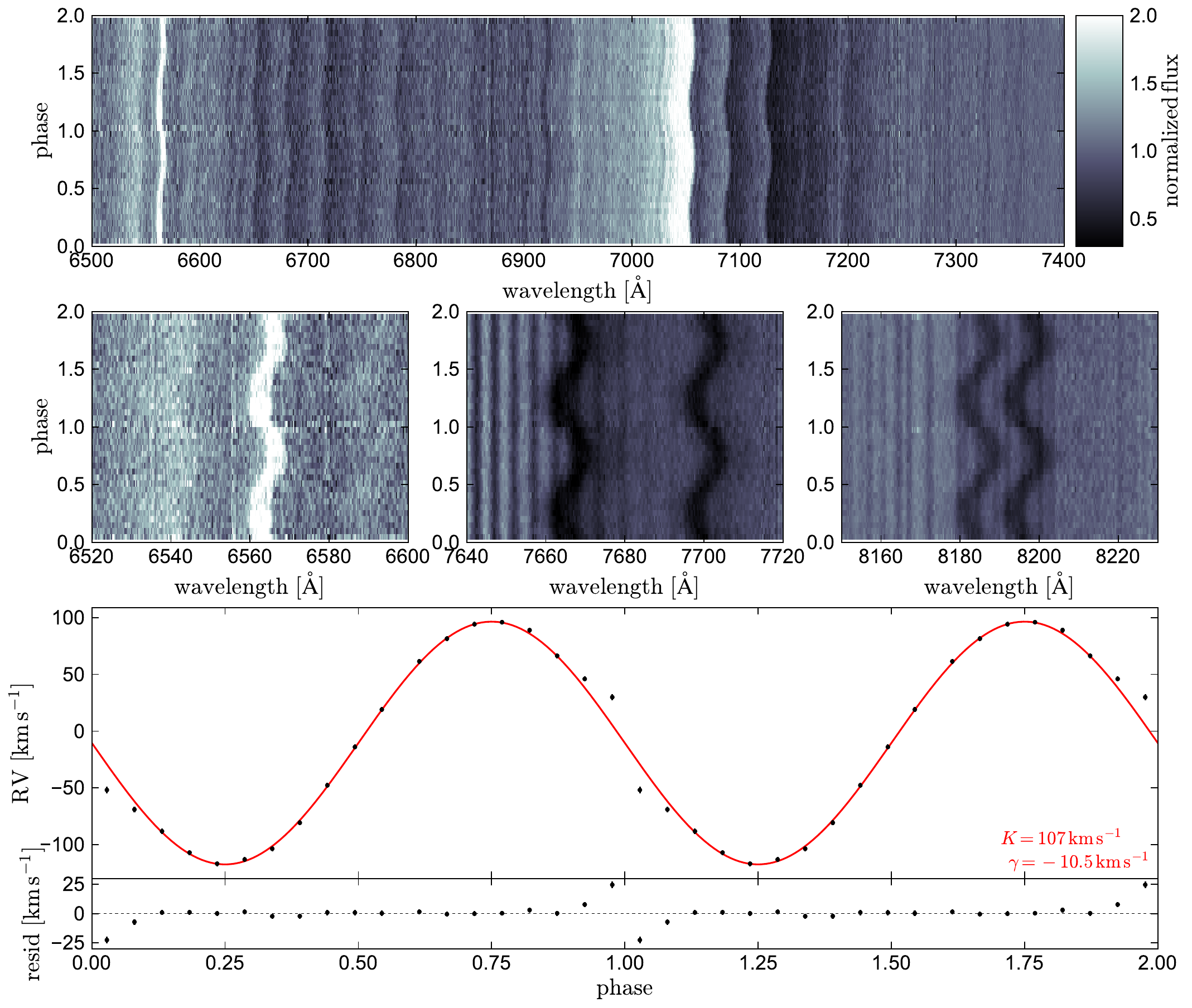}
    \caption{Phased ESI spectra of \ztfb. Top panel shows a broad window spanning several orders; middle panels show cutouts of spectral regions including the H$\alpha$ line (left), the K\,I resonance doublet (middle) and the Na I doublet (right). In all panels, spectra are phased according to the photometric ephemeris and duplicated once. Stationary absorption lines are tellurics. There is no evidence of a second component moving in anti-phase of the primary.  Bottom panel shows the measured RVs and best-fit sinusoid. The RVs are well-described by a sinusoid except during the primary eclipse (phase 1.0), when a clear Rossiter-McLaughlin effect is visible.   }
    \label{fig:river}
\end{figure*}

Most of the lines in the observed spectrum are broad and not well suited for measuring RVs. After some experimentation, we chose two separate spectral windows for RV measurements: the region between 8160 and 8220\,\AA, which includes the  Na I $\lambda \lambda$8183, 8195 doublet, and the region between 7655 and 7725\,\AA, which includes the  K I $\lambda \lambda$7665, 7699 doublet. These two regions are shown in the center middle and right panels of Figure~\ref{fig:river}. We pseudo-continuum normalized both the observed and model spectra by dividing by a running median in a 100\,\AA\, window. The Na I doublet appears in two separate ESI orders, which we analyze separately. For each window and exposure, we determine the RV that minimizes $\chi^2$, and estimate the uncertainty as the RV shift that corresponds to a $\chi^2$ increase of 1. Finally, we report the inverse variance-weighted average of the RVs from the three windows. The RVs are reported in Table~\ref{tab:obslog}.

The bottom panel of Figure~\ref{fig:river} shows the measured RVs. At most phases, they are well-described by a sinusoid with RV semi-amplitude $K=107.8\pm 1.2\,\rm km\,s^{-1}$, corresponding to a mass function $f(M_2) = 0.0103\pm 0.0003\,M_{\odot}$. However, there is a clear discontinuity at phase 1, which is also visible in the trailed spectra. This is a Rossiter-McLaughlin (RM) effect \citep{Rossiter1924, McLaughlin1924}, reflecting the fact that the BD first eclipses the blueshifted side of the rotating M dwarf, and later the redshifted side.

\subsubsection{Projected rotation velocity}
\label{sec:vsini}
We constrain the projected rotation velocity, $v\sin i$, of the M dwarf from the broadening of the Na I $\lambda \lambda$8183, 8195 doublet. To minimize orbital smearing and biases resulting from the RM effect, we only consider the 7 ESI exposures taken near conjunction (i.e., $\left|\phi-0.25\right|<0.09$ and $\left|\phi-0.75\right|<0.09$). We verified using simulations that the expected bias in $v\sin i$ due to orbital smearing in these exposures is $\lesssim 3\,\rm km\,s^{-1}$. We corrected these exposures for telluric absorption, co-added them in their rest frame, and compared the resulting spectrum to a grid of \texttt{BT-Settl} models with a range of $v\sin i$ values. This yielded a projected rotation velocity $v \sin i= 118\pm 6\,\rm km\,s^{-1}$ for the M dwarf. 

Our joint modeling of the light curve, RVs, and SED (Section~\ref{sec:jointmodel}) predicts $v\sin i = 2\pi R_\star \sin i /P_{\rm orb} = 111.7\pm 1.2\,\rm km\,s^{-1}$ if the M dwarf is tidally synchronized, consistent with this measurement.

\subsubsection{Spectral type}
\label{sec:sptype}

We compared the flux-calibrated, coadded ESI spectrum to a grid of empirical spectral models created by \citet{Kesseli2017}. That work produced high-SNR templates for each spectral subtype in the MK system by coadding SDSS/BOSS spectra of bright stars with that subtype. Spectral subtypes for stars in the library were assigned using the automated ``Hammer'' classification scheme described by \citet{Covey2007}.

\begin{figure*}
    \centering
    \includegraphics[width=\textwidth]{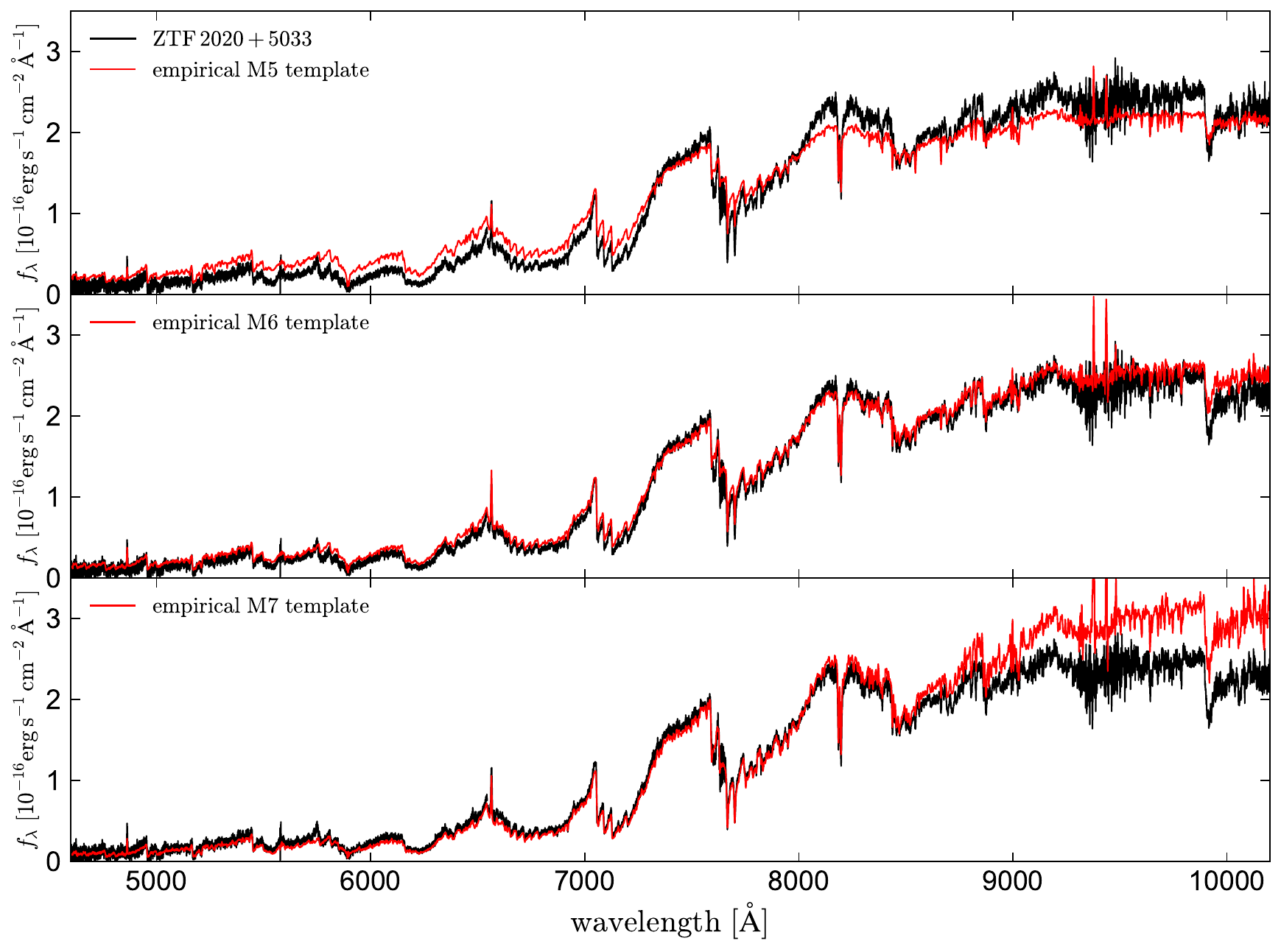}
    \caption{Coadded, rest-frame spectrum of \ztfbd (black) compared to empirical templates for M dwarfs with a range of spectral types (red). All templates are scaled to match the observed flux at 8000\,\AA. The M5 template is bluer and has shallower spectra lines than the data, while the M7 template is redder. The M6 template provides the best match.  }
    \label{fig:sptyep}
\end{figure*}

We compare the three closest-matching templates to the coadded rest-frame spectrum of \ztfbd in Figure~\ref{fig:sptyep}. As expected, the best-fitting templates are for dwarf stars with mid- to late-M spectral types. The M5 template clearly has an earlier spectral type than \ztfbd: it is bluer and has shallower absorption features at red wavelengths. The M6 template is a reasonably good match in both broad spectral shape and in the strength of individual lines, though it slightly overpredicts the flux at bluer wavelengths. The M7 template is redder than the observed spectrum. 

\subsection{Spectral energy distribution}
\label{sec:SED}

We constructed the broadband spectral energy distribution (SED) of \ztfbd by combining photometry from the Pan-STARRS \citep[DR2;][]{Kaiser2002}, 2MASS \citep{Skrutskie2006}, and WISE \citep{Wright2010} surveys. The Pan-STARRS and WISE magnitudes are mean values calculated from 10+ epochs and thus represent the source's time-averaged flux. The 2MASS magnitudes were measured in a single exposure at JD 2450994.9554, 24 years prior to $t_0$, making it impractical to measure their phases reliably in the presence of period modulations (Section~\ref{sec:ttvs}). The fact that our SED model reproduces the photometry from all surveys without significant systematics suggests that the 2MASS observations did not occur during the primary eclipse. The SED and best-fit model are shown in the right panel of Figure~\ref{fig:summary}; mock photometry for both components is shown in the bottom right panel of Figure~\ref{fig:fitting_results}.

\subsubsection{Primary mass from $K_s$-band absolute magnitude}
\label{sec:Mks}
The measured $K_s$-band magnitude of the binary is $K_s=14.33\pm 0.06$, corresponding to an absolute magnitude $M_{K_s} = 8.56 \pm 0.07$ at the distance inferred from the {\it Gaia} parallax. Our light curve fit implies that the BD contributes about 10\% of the flux in the $K_s$-band, so the M dwarf has an absolute magnitude $M_{K_s} = 8.66 \pm 0.07$. The empirical $M_{K_s} - M_\star$ relation constructed by \citet{Mann2019} then predicts $M_\star = 0.132 \pm 0.005 \,M_{\odot}$, where the uncertainty accounts for both intrinsic scatter in the relation and uncertainty in $M_{K_s}$. This mass is on the high end of values predicted for an M6 spectral type \citep[e.g.][]{Baraffe1996}, which range from 0.10 to 0.13\,$M_{\odot}$, depending on the adopted conversion between color and spectral type.  

\subsubsection{Metallicty}
\label{sec:feh}
We estimated the metallicity of the M dwarf using the empirical calibration from \citet{Mann2013}. In brief, we measured equivalent widths of several temperature- and metallicity-sensitive spectral features in the coadded and flux-calibrated ESI spectrum, defining the local pseudo-continuum for each feature with a first-order polynomial fit to predetermined regions on either side of each feature. \citet{Mann2013} calculated an empirical metallicity estimator as a function of these features (their Equation 11) using M dwarfs in wide binaries with FGK companions of known metallicity. Their relation yields $\rm [M/H]=-0.44$ for \ztfb, with an estimated uncertainty of $\sim 0.1$ dex.

\subsection{Galactic orbit and kinematic age}
\label{sec:Gal_orbit}
\ztfbd has a tangential velocity relative to the Sun of $v_\perp = 4.74\mu /\varpi = 98\,\rm km\,s^{-1}$, suggesting membership in an old and kinematically hot population. We calculated a Galactic orbit for the source using \texttt{galpy} \citep[][]{Bovy2015}. We used the parallax and proper motion from {\it Gaia} DR3 \citep{GaiaCollaboration2021, GaiaCollaboration2022}, together with the center-of-mass RV inferred from the joint fit, as starting points to compute its orbit backward in time for 500 Myr. We used the Milky Way potential from \citet{McMillan2017}, but our results are only weakly sensitive to this choice. The orbit is characteristic of the thick disk, with moderate eccentricity and maximum excursions of $\pm$600 pc from the midplane. The current Galactocentric velocities in a cylindrical frame are $(v_R, v_\phi, v_z) = (-102, -228, +27)\,\rm km\,s^{-1}$. These velocities, particularly the large $v_R$, are significantly different from the local standard of rest and point towards the source's membership in a kinematically hot population.

While Galactic kinematics provide only rough constraints on stellar ages, the orbit of \ztfbd rules out a young age. In the solar neighborhood, essentially no stars with ages below 5 Gyr have velocities in excess of 100 $\rm km\,s^{-1}$ with respect to the Sun \citep[e.g.][]{Seabroke2007, Yu2018}. Even among stars older than 10 Gyr, a majority are slower than \ztfbd \citep{Sharma2014}. We adopt a conservative lower limit of 5 Gyr for the system's age.

\begin{table*}
\centering
\caption{Physical parameters and 1$\sigma$ uncertainties for both components of \ztfb. }
\begin{tabular}{lll}
\hline\hline
\multicolumn{3}{l}{\bf{Observables of the unresolved source}}  \\ 
Right ascension [J2016.0] & $\alpha$\,[deg] & 305.014397 \\
Declination  [J2016.0] & $\delta$\,[deg] & 50.560459 \\
Apparent magnitude & $G$\,[mag] & 18.70 \\
Parallax & $\varpi$\,[mas] & $7.05\pm 0.15$ \\
Proper motion (RA) & $\mu_\alpha ^*$\,[$\rm mas\,yr^{-1}$] & $-105.02\pm 0.17$ \\
Proper motion (Dec) & $\mu_\delta$\,[$\rm mas\,yr^{-1}$] & $-100.21\pm 0.19$ \\
Tangential velocity & $v_\perp$\,[$\rm km\,s^{-1}$] & $97.6 \pm 2.1$ \\
Extinction & $E(B-V)$ [mag] & 0.0 \\
\hline

\multicolumn{3}{l}{\bf{Parameters of the M dwarf}}  \\ 
Effective temperature & $T_{\rm eff, \star}$\,[K] & $2856 \pm 6$ \\
Radius & $R_\star\,[R_{\odot}]$ & $0.176\pm 0.002$  \\ 
Mass &  $M_\star\,[M_{\odot}]$ & $0.134 \pm 0.004$ \\ 
Projected rotation velocity & $v_\star\sin i$\,[km\,s$^{-1}$] &  $118\pm 6$ \\
Surface gravity   & $\log(g_\star/(\rm cm\,s^{-2}))$  & $5.07\pm0.01$  \\
Bolometric luminosity & $\log(L_\star/L_{\odot})$ & $-2.73 \pm 0.01$ \\

\hline
\multicolumn{3}{l}{\bf{Parameters of the brown dwarf}}   \\ 
Effective temperature & $T_{\rm eff,\rm BD}$\,[K] & $1691 \pm 127$ \\
Mass &  $M_{\rm BD}\,[M_{\rm J}]$ & $80.1\pm 1.6$ \\ 
Radius & $R_{\rm BD}\,[R_{\rm J}]$ & $1.05\pm 0.01$  \\ 
Surface gravity   & $\log(g_{\rm BD}/(\rm cm\,s^{-2}))$  & $5.25\pm 0.01$  \\
Bolometric luminosity & $\log(L_{\rm BD}/L_{\odot})$ & $ -4.07\pm 0.12$ \\ 
Reflection parameter & $\alpha_{\rm BD}$ & $0.14 \pm 0.07$ \\ 

\hline
\multicolumn{3}{l}{\bf{Parameters of the binary}}   \\ 
Orbital period & $P$\,[day]  & $0.07928432 \pm 0.000000016$  \\
Conjunction time & $t_0$\,[HJD UTC]  & $2459733.9125 \pm 0.0002$  \\

M dwarf RV semi-amplitude  & $K_{\star}$ [km\,s$^{-1}$] & 107.8$\pm$1.2 \\
M dwarf center-of-mass velocity & $\gamma_{\star}$\,[km\,s$^{-1}$] & $-10.1 \pm 0.6$ \\ 
Mass ratio  &  $q = M_{\rm BD}/M_{\star}$  & $0.571\pm 0.009$ \\
Orbital inclination & $i\,[\rm deg]$ & $87.6\pm 1.6$  \\
Semimajor axis  & $a$ [R$_{\odot}$]   &  $0.462\pm 0.004$ \\
Distance  & $d$ [pc]   &  $136.4\pm 1.9$ \\
Age  & $\tau$ [Gyr]   &  $5-13$ \\
Gravitational wave inspiral time  & $t_{\rm GW}$ [Gyr]   &  $1.3\pm 0.1$ \\
\hline

\hline
\end{tabular}
\begin{flushleft}

\label{tab:system}
\end{flushleft}
\end{table*}

\begin{table}
\begin{tabular}{llll}
JD UTC & phase &  RV\,[$\rm km\,s^{-1}$]  & SNR  \\
\hline
2459734.0405 & 0.61 & $61.5 \pm 2.2$ & 14.7 \\
2459734.0446 & 0.67 & $81.5 \pm 2.2$ & 14.5 \\
2459734.0487 & 0.72 & $94.2 \pm 2.2$ & 15.1 \\
2459734.0528 & 0.77 & $96.0 \pm 2.1$ & 15.0 \\
2459734.0569 & 0.82 & $89.0 \pm 2.2$ & 15.0 \\
2459734.0610 & 0.87 & $66.4 \pm 2.2$ & 14.3 \\
2459734.0651 & 0.92 & $46.0 \pm 2.2$ & 13.5 \\
2459734.0692 & 0.98 & $29.9 \pm 2.8$ & 10.9 \\
2459734.0733 & 0.03 & $-51.9 \pm 2.8$ & 11.0 \\
2459734.0774 & 0.08 & $-69.1 \pm 2.3$ & 14.1 \\
2459734.0815 & 0.13 & $-88.3 \pm 2.2$ & 15.4 \\
2459734.0856 & 0.18 & $-107.1 \pm 2.3$ & 15.3 \\
2459734.0897 & 0.23 & $-116.9 \pm 2.2$ & 15.6 \\
2459734.0938 & 0.29 & $-113.2 \pm 2.3$ & 16.0 \\
2459734.0979 & 0.34 & $-103.7 \pm 2.2$ & 15.4 \\
2459734.1020 & 0.39 & $-80.8 \pm 2.2$ & 15.1 \\
2459734.1061 & 0.44 & $-47.8 \pm 2.3$ & 14.6 \\
2459734.1102 & 0.49 & $-14.0 \pm 2.3$ & 14.2 \\
2459734.1142 & 0.54 & $19.0 \pm 2.2$ & 12.5 \\

\end{tabular}
\caption{Radial velocities from the ESI spectra. Timestamps are calculated at mid-exposure, and phases are calculated from the photometric ephemeris. SNR is calculated at 8200\,\AA.}
\label{tab:obslog}
\end{table}

\begin{table}
\begin{tabular}{cccc}
Filter & system & central wavelength\,[$\mu$\,m] & mag  \\
\hline
$g$ & AB & 0.48 & $21.27 \pm 0.03$ \\
$r$ & AB & 0.62 & $19.90 \pm 0.06$ \\
$i$ & AB & 0.75 & $17.95 \pm 0.03$ \\
$z$ & AB & 0.87 & $17.03 \pm 0.03$ \\
$y$ & AB & 0.96 & $16.56 \pm 0.03$ \\
$J$ & Vega & 1.24 & $15.15 \pm 0.05$ \\
$H$ & Vega & 1.65 & $14.58 \pm 0.04$ \\
$K_s$ & Vega & 2.17 & $14.33 \pm 0.06$ \\
$W_1$ & Vega & 3.35 & $14.00 \pm 0.05$ \\
$W_2$ & Vega & 4.60 & $13.84 \pm 0.05$ \\

\end{tabular}
\caption{Spectral energy distribution. We adopt an uncertainty floor of 0.03 mag in all filters.}
\label{tab:SED}
\end{table}

\section{Parameter inference}
\label{sec:jointmodel}
We constrain the masses, radii, and temperatures of both components, as well as the binary's orbital inclination, with a joint fit of the measured RVs, CHIMERA and {\it WISE} light curves, the {\it Gaia} parallax, and the broadband SED. The role of each observable in constraining the system parameters is complex given the several overlapping constraints, but we summarize the most important constraints as follows. 
\begin{enumerate}
    \item The temperature and radius of the M dwarf are constrained by the SED and ${\it Gaia}$ parallax. 
    \item The density of the M dwarf is constrained by the amplitude of ellipsoidal variations in the light curve, which are the result of its tidal deformation. Given the radius constraint from (1), this constrains the M dwarf's mass.
    \item The mass of the BD is constrained by the amplitude of the M dwarf's RV variability, with a weak dependence on inclination.
    \item The radius of the BD and the inclination are both constrained by the primary eclipse and by the observed RM effect. 
    \item The temperature of the BD is constrained by the depth of the secondary eclipse and its dependence on wavelength.  
\end{enumerate}

\subsection{Light curve and RV models}
\label{sec:lc_model}
We model the RVs and light curves in the $g-$, $z_s$, and $W1$ bands using \texttt{ellc} \citep{Maxted2016}, a flexible code for calculating light curves and flux-weighted RVs of detached binaries. We use \texttt{ellc} rather than alternatives such as \texttt{PHOEBE} \citep{Prsa2005} because it allows us to specify the surface brightness ratio -- which we calculate using  \texttt{BT-Settl} model SEDs --  in each bandpass, which is more accurate than e.g. modeling the BD as a blackbody. Both components are modeled using the code's \texttt{Roche} geometry. We take limb darkening and gravity darkening coefficients for each bandpass from \citet{Claret2012}, using their 4-term parameterization for the limb darkening law. These authors did not calculate coefficients for the {\it WISE} $W1$ band, so we use their coefficients for the similar {\it Spitzer}/IRAC 3.6\,$\mu$m band. We model reflection (i.e., heating of the BD's ``day'' side by the M dwarf) using the simplified model described by \citet{Maxted2016}, leaving the albedo coefficient as a free parameter of the fit. 

We calculate the surface brightness ratio of the two components in each band by integrating \texttt{BT-Settl} model spectra over the bandpass with \texttt{pyphot}.\footnote{https://mfouesneau.github.io/pyphot/} We use the same process to calculate phase-averaged mean magnitudes of both components for comparison with the observed SED. 

To model the RVs and RM effect, we first calculate the instantaneous RV curve, with the RV of each element of the M dwarf's surface weighted by that element's flux at 8200\,\AA. To account for finite exposure times, we then calculate a flux-weighted average over the 300-second exposure time of the ESI spectra.

\subsection{Joint fit}
\label{sec:likelihood}

We use a likelihood function that combines constraints from RVs, light curves, the SED, and astrometry:
\begin{equation}
    \label{eq:lnN}
    \ln L=\ln L_{{\rm RVs}}+\ln L_{{\rm light\,curves}}+\ln L_{{\rm SED}}+\ln L_{{\rm ast}}. 
\end{equation}
Here $\ln L_{\rm RVs}$ compares the predicted and measured RVs:
\begin{equation}
    \label{eq:lnL_RVs}
    \ln L_{{\rm RVs}}=-\frac{1}{2}\sum_{i}^{N_{\rm RVs}}\frac{\left({\rm RV}_{{\rm pred}}\left(t_{i}\right)-{\rm RV}_{i}\right)^{2}}{\sigma_{{\rm RV,}i}^{2}},
\end{equation}
where ${\rm RV}_i$ are the RVs measured at times $t_i$, $\sigma_{\rm RV,i}$ are their uncertainties, and ${\rm RV}_{{\rm pred}}\left(t_{i}\right)$ are the exposure-averaged RVs predicted by \texttt{ellc}. 

Similarly, $\ln L_{\rm light\,curves}$ compares the predicted and observed light curves:

\begin{equation}
    \label{eq:lnL_light_curves}
    \ln L_{{\rm light\,curves}}=-\frac{1}{2}\sum_{\lambda}^{N_{{\rm bands}}}\sum_{i}^{N_{{\rm epochs}}}\frac{\left(\hat{f}_{{\rm \lambda\,pred}}\left(t_{i}\right)-\hat{f}_{\lambda,i}\right)^{2}}{\sigma_{\hat{f_{\lambda}},i}^{2}},
\end{equation}
where $\hat{f}_{\lambda,i}$ represents the $i$th measurement of normalized flux in the $\lambda$th band, $\sigma_{\hat{f_{\lambda}},i}$ is its uncertainty, and $\hat{f}_{{\rm \lambda\,pred}}\left(t_{i}\right)$ is the corresponding quantity predicted by \texttt{ellc}. The inner summation is over all photometric points in a given light curve and the outer summation is over the three different bandpasses. 

 $\ln L_{\rm SED}$ compares the observed and predicted SED:

\begin{equation}
    \label{eq:lnL_SED}
    \ln L_{{\rm SED}}=-\frac{1}{2}\sum_{\lambda}\frac{\left(m_{{\rm pred,}\lambda}-m_{{\rm \lambda}}\right)^{2}}{\sigma_{m_{\lambda}}^{2}},
\end{equation}
where $m_\lambda$ and $\sigma_{m_{\lambda}}$ are the observed magnitudes of the source in different bandpasses and their corresponding uncertainties, and $m_{{\rm pred,}\lambda}$ are the predicted magnitudes of the unresolved source in each bandpass. We calculate $m_{{\rm pred,}\lambda}$ by interpolating on a grid of \texttt{BT-Settl} model spectra, summing the predicted flux from the M dwarf and brown dwarf, and using \texttt{pyphot} to compute bandpass-averaged magnitudes.

\begin{figure*}
    \centering
    \includegraphics[width=\textwidth]{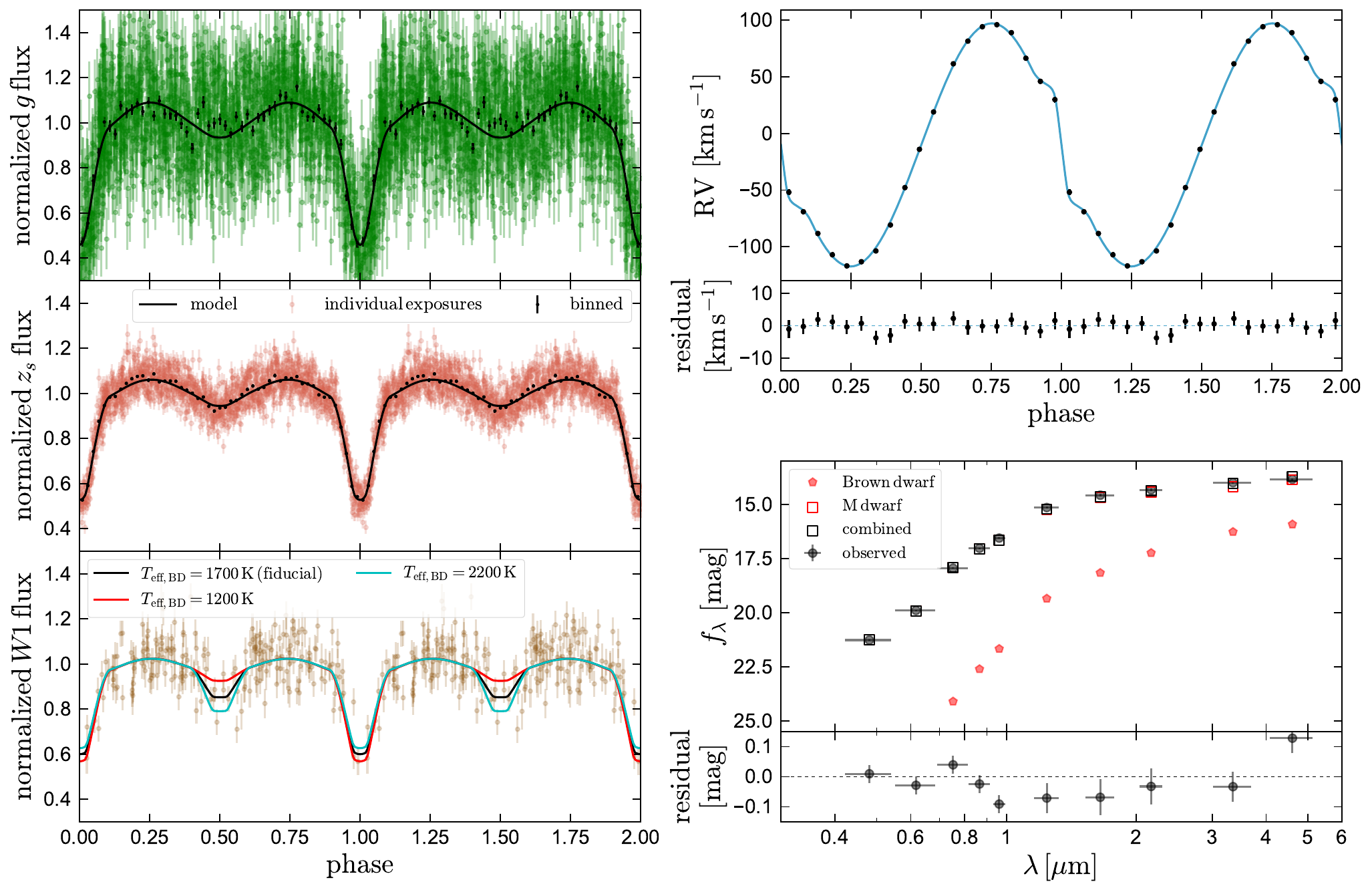}
    \caption{Posterior predictive checks for joint fitting of light curves, RVs, and the SED. Left panel shows the best-fit \texttt{ellc} light curves, compared to the observed light curves in the $g$, $z_s$, and $W1$ bands. Black points in the upper left and center panels show binned averages of the data. Bottom left panel shows how the secondary eclipse in the $W1$ band constrains the temperature of the BD. Upper right panel shows the predicted instantaneous RV curve; the residuals account for the finite exposure time of the observed RVs. Bottom right shows the fit to the SED, with the contributions from the two components shown separately. The fit is successful in matching all observables without obvious systematics. }
    \label{fig:fitting_results}
\end{figure*}

Finally, $\ln L_{\rm ast}$ compares the observed and predicted parallaxes:
\begin{equation}
    \ln L_{{\rm ast}}=-\frac{1}{2}\frac{\left(1/d-\varpi\right)^{2}}{\sigma_{\varpi}^{2}},
\end{equation}
where $d$ is the distance in kpc, $\varpi$ is is the zeropoint-corrected {\it Gaia} parallax, and $\sigma_{\varpi}$ is its uncertainty.

We adopt flat priors on all parameters except the M dwarf's mass, for which we use a Gaussian prior $M_\star\,[M_{\odot}] \sim \mathcal{N}(0.0132, 0.005)$ that is motivated by the empirical $M_{K_s} - M_\star$ relation (Section~\ref{sec:Mks}). We sample from the posterior using \texttt{emcee} \citep{Foreman-Mackey2013} using 64 walkers and taking 1000 steps after an initial 1000-step burn-in period.

\subsection{Results}

\begin{figure*}
    \centering
    \includegraphics[width=\textwidth]{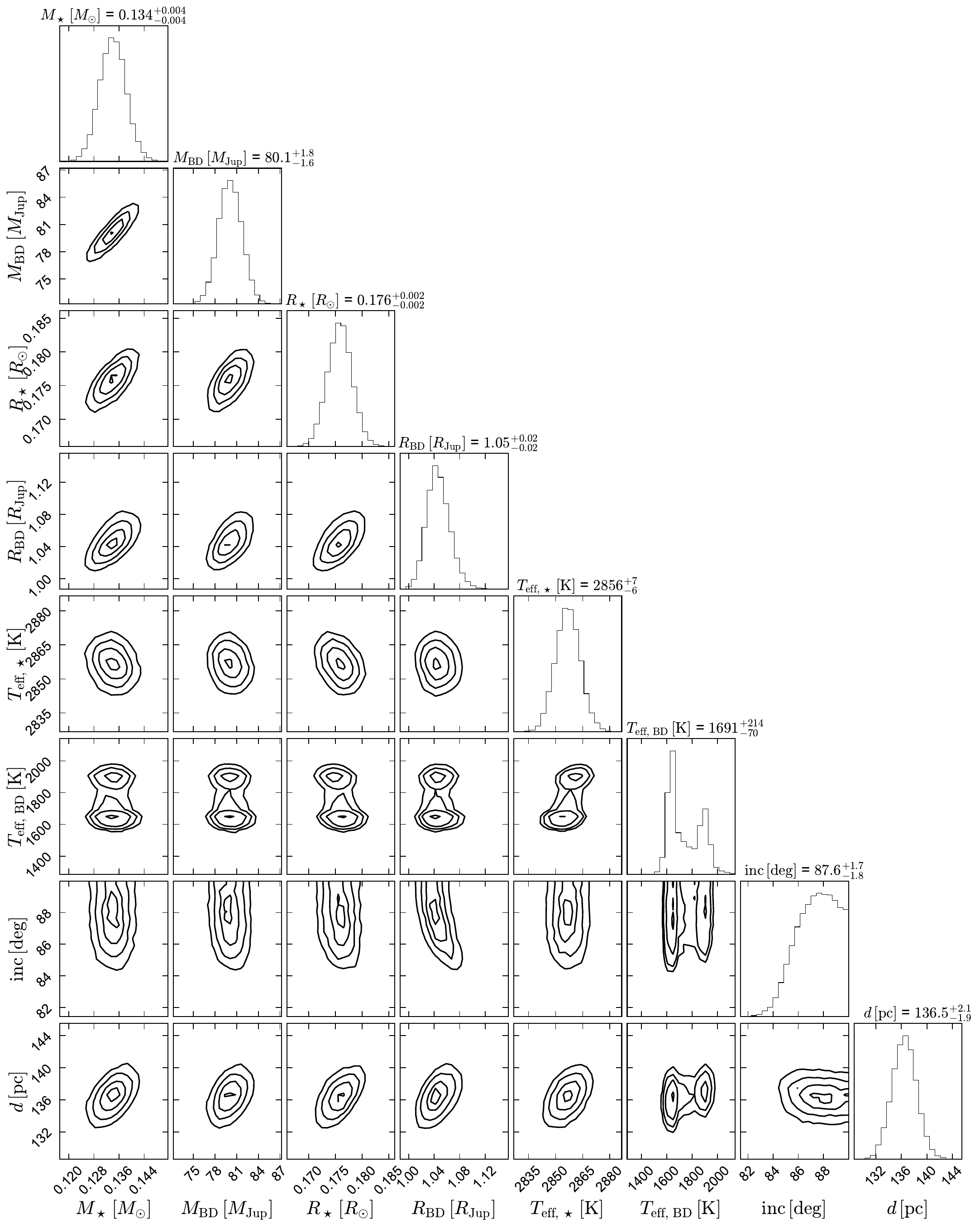}
    \caption{Constraints from joint fitting of light curves, RVs, and the broadband SED. Parameters with a ``$\star$'' subscript refer to the M dwarf, while those with the subscript ``BD'' refer to the brown dwarf. Masses, radii, and temperatures of both components are constrained with a precision of a few percent or better (see Table~\ref{tab:system} for constraints on all parameters).}
    \label{fig:corner}
\end{figure*}

Figure~\ref{fig:fitting_results} compares the predicted light curves, RVs, and SED to the observed data. The fit is in general quite good, with no obvious systematic differences between the model predictions and data. Joint constraints on several of the free parameters are shown in Figure~\ref{fig:corner} and reported in Table~\ref{tab:system}. Several parameters are covariant, as expected: 
\begin{itemize}
    \item $R_\star$ and $d$ are covariant because both rescale the predicted SED. 
    \item $R_\star$ and $M_\star$ are covariant because the observed ellipsoidal variation constrains $M_\star/R_\star ^3$.
    \item $M_\star$ and $M_{\rm BD}$ are covariant because the M dwarf's predicted RVs increase with $M_{\rm BD}$ and decrease with $M_\star$. 
    \item $R_{\rm BD}$ and $R_{\star}$ are covariant because their ratio sets the primary eclipse depth.
\end{itemize}
These joint covariances also lead to covariances between parameters that are not obviously physically covariant, such as $M_{\rm BD}$ and $R_\star$. The constraint on the temperature of the BD is bimodal, with a dominant mode at $T_{\rm eff,\,BD}\approx 1600$\,K and a second mode at $T_{\rm eff,\,BD}\approx 1850$\,K. This occurs because the \texttt{BT-Settl} model spectra predict the $z_s$-band flux to vary non-monotonically with $T_{\rm eff}$, with slightly lower predicted flux (and thus, a shallower secondary eclipse) at $1700\,\rm K$ than at 1600 or 1800\,\rm K.

The best-fit mass of the M dwarf is $M_\star = 0.134\pm 0.004\,M_{\odot}$, which is 0.5$\sigma$ above the prior from the \citet{Mann2019} relation. Since we have not measured RVs for the BD, the main constraint on $M_\star$ comes from the amplitude of ellipsoidal variation in the light curve, and from the prior. The inferred mass of the BD is $M_{\rm BD} = 80.1\pm 1.6\,M_{\rm J}$, almost exactly equal to the theoretically expected hydrogen burning limit. This result is only weakly sensitive to varying the mass of the M dwarf within its uncertainties: if the mass of the M dwarf were $0.132\,M_{\odot}$ (the central value of the prior as informed by the $K_s-$ band absolute magnitude), the RVs would imply $M_{\rm BD}= 79.2\,M_{\rm J}$ -- still very close to the hydrogen burning limit. 

The boundary between BDs and low-mass stars is near  $\approx 80\,M_{\rm J}$, with some dependence on metallicity \citep{Burrows1993, Baraffe2002, Dieterich2014, Baraffe2015}, so our inferred $M_{\rm BD}$ is consistent with either a high-mass BD or a very low-mass star. However, the inferred temperature of the BD, $T_{\rm eff, BD}=1691\pm 126\,\rm K$, is cooler than both theoretical (Section~\ref{sec:evol_models}) and empirical \citep[e.g.][]{Dieterich2014} constraints on the minimum temperature of main-sequence stars, leading us to interpret the object as a probable BD.

\section{Discussion}
\label{sec:discussion}

\begin{figure}
    \centering
    \includegraphics[width=\columnwidth]{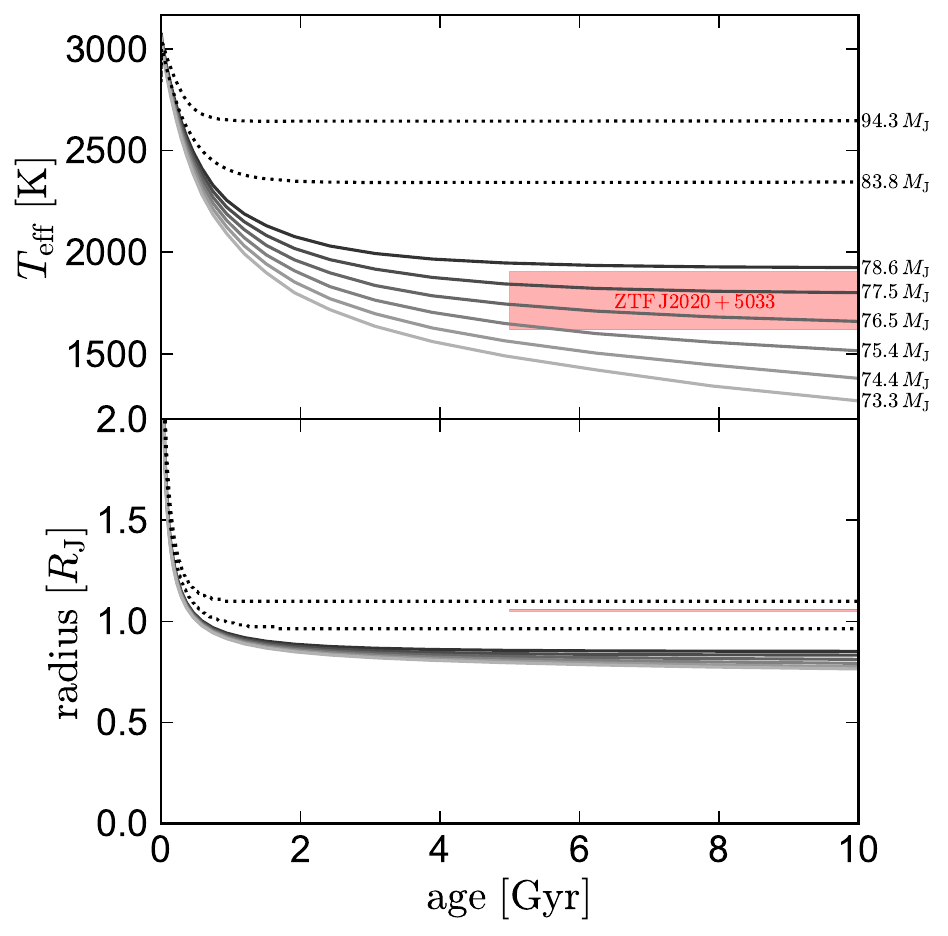}
    \caption{Temperature and radius of \ztfbd compared to evolutionary models for massive brown dwarfs (solid lines) and very low-mass stars (dotted lines). Given the system's old age (as constrained by its Galactic orbit), its effective temperature implies a mass of $76-78\,M_{\rm J}$. This is quite close to the dynamically-inferred mass of $80.1\pm 1.7\,M_{\rm J}$, and {\it just} at the hydrogen burning limit. The measured radius is $\approx 25\%$ larger than predicted by evolutionary models.  }
    \label{fig:evols}
\end{figure}

\begin{figure*}
    \centering
    \includegraphics[width=\textwidth]{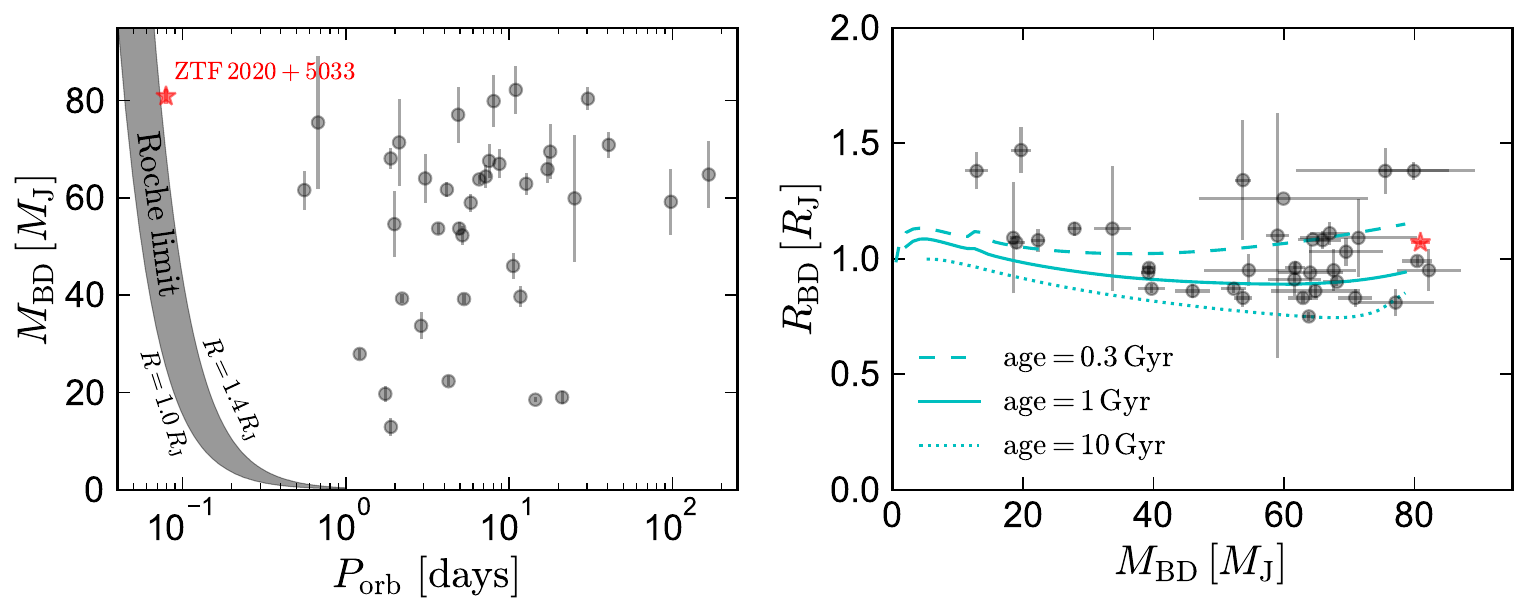}
    \caption{Comparison of \ztfbd to other transiting BDs, as compiled by \citet{Carmichael2023}. Left: \ztfbd has by far the shortest orbital period of known BDs orbiting stars, and one of the highest masses. BDs below the shaded region would overflow their Roche lobes; \ztfbd is the only object near this limit. Right: masses and radii for the same systems shown in the left panel. Cyan lines show predictions from ATMO 2020 evolutionary models \citep{Phillips2020} for three different ages. \ztfbd has a radius similar to other observed high-mass BDs. However, its radius would be consistent with the ATMO 2020 models only at an age of order 400 Myr, which is much younger than implied by its Galactic orbit. This suggests the brown dwarf is somewhat inflated.  }
    \label{fig:pop}
\end{figure*}

\subsection{Comparison to evolutionary models}
\label{sec:evol_models}
Figure~\ref{fig:evols} compares our constraints on the temperature and radius of the BD to solar-metallicity ATMO 2020 evolutionary models for BDs \citep{Phillips2020} and very low-mass star models from \citet{Baraffe2015}. We assume a minimum age of 5 Gyr, as motivated by the star's thick disk-like Galactic orbit (Section~\ref{sec:Gal_orbit}). At this age, the observed temperature is consistent with the highest-mass BD models in the ATMO 2020 library, which have masses between 75 and 79\,$M_{\rm J}$. This is consistent with our dynamically-inferred mass. 

On the other hand, the BD's radius is significantly larger than the models predict. The 78.6\,$M_{\rm J}$ model predicts a radius of $0.85\,R_{\rm J}$ at ages of $>5\,$Gyr, meaning that at $1.06\,R_{\rm J}$, the BD is $\approx$25\% larger than expected. The model matches the observed radius only at an age of 400\,Myr, which is implausibly young given the system's kinematics (Section~\ref{sec:Gal_orbit}). This implies that the BD is inflated. 

We also compared to the BD's parameters to the Sonora evolutionary models from \citet{Marley2021}, assuming a metallicity $\rm [Fe/H]=-0.5$. This yielded results quite similar to those from the ATMO 2020 models: the BD's temperature implies a mass of $76-79\,M_{\rm J}$ for ages of 5-10 Gyr, and the predicted radius of a non-inflated BD in this age range is $0.80-0.86\,R_{\rm J}$. 

The M dwarf is also somewhat larger than predicted by evolutionary models, with a radius of $0.175\pm 0.002\,R_{\odot}$ at a mass where evolutionary models \citep[e.g.][]{Baraffe2015} predict a radius of $\approx 0.16\,R_{\odot}$. This radius is, however, typical of observed M dwarfs in this mass range \citep[e.g.][]{Parsons2018}.

\subsection{Is the secondary a BD or a low-mass star?}
\label{sec:BD_or_star}
The secondary mass we infer is very close to the hydrogen burning limit, raising the possibility that the object could be a low-mass star rather than a BD. Recent evolutionary calculations place the hydrogen burning limit at $\approx 78.5\,M_{\rm J}$ at solar metallicity \citep{Chabrier2023}, rising toward lower metallicities. The exact value of the limit is uncertain at the $\sim 0.02\,M_{\rm J}$ level due to uncertainties in evolutionary models, particularly the equation of state. We thus cannot distinguish between star and BD secondaries on the basis of mass alone.

Even at old ages, models predict a smooth transition between the lowest-mass stars and highest-mass BDs, with the highest-mass BDs undergoing unsteady hydrogen burning that contributes a significant fraction of their luminosity even after 10 Gyr \citep{Chabrier1997, Zhang2017}. We classify the secondary as a BD because it is cooler than the stellar/substellar boundary at $\sim$2100\,K empirically-inferred by \citet{Dieterich2014}. Given the uncertainty in this limit and  \citep[e.g.][]{Dupuy2017}, we cannot fully rule out the possibility that the secondary is a low-mass star, but we consider a BD more likely.

\subsection{Comparison to other known BDs}
\label{sec:population}

Figure~\ref{fig:pop} compares \ztfbd to 39 other transiting BDs whose masses, radii, and orbital periods were tabulated by \citet{Carmichael2023}. These systems represent a majority of the currently known transiting BD population. All of these host stars are the main sequence; BDs orbiting white dwarfs were not included. The most unusual property of the system compared to the known population is its orbital period, which is 7 times shorter than the previous record holder, TOI 263.01 \citep{Parviainen2020, Palle2021}. The shaded region in the left panel shows the boundary below which BDs with radii between 1.0 and 1.4\,$R_{\rm J}$ would overflow their Roche lobes. The BD in \ztfbd still falls comfortably within its Roche lobe ($R_{\rm BD}/R_{\rm Roche\,lobe} \approx 0.71$) but would overflow it if its radius were larger than $1.5\,R_{\rm J}$, or at its current radius if its mass were lower than $30\,M_{\rm J}$.

Most of the other BDs shown in Figure~\ref{fig:pop} transit solar-type stars, which are less dense than the M dwarf in \ztfb. Solar-type stars with $80\,M_{\rm J}$ companions overflow their Roche lobes at $P_{\rm orb} \lesssim 6$ hours (0.25 days). A period as short 1.9 hours can only occur around a low-mass main sequence star, BD, or compact object. Nevertheless, none of the other BDs shown in Figure~\ref{fig:pop} orbit stars that fill or nearly fill their Roche lobes, so the dearth of BDs with $P_{\rm orb} \lesssim 1\,\rm d$ is not (purely) a consequence of the Roche limit. 

The secondary in \ztfbd is among the most massive known BDs, assuming it indeed is a BD. The right panel of Figure~\ref{fig:pop} shows that the object's radius is typical for the population of observed high-mass BDs. There are two other objects with similar masses and larger radii: TOI-587b, with $M=79.9 \pm 5.3\,M_{\rm J}$ and $R = 1.38 \pm  0.04\,R_{\rm J}$ \citep{Grieves2021}, and NGTS-7Ab, with $M=75.5 \pm 13.7\,M_{\rm J}$ and $R = 1.38 \pm  0.04\,R_{\rm J}$ \citep{Jackman2019}. However, both of these systems are known to be young, with respective inferred ages of 200 and 55 Myr. The object most similar to \ztfbd is thus KOI-189b,  with $M=80.4 \pm  2.3\,M_{\rm J}$ and $R = 0.99 \pm  0.02\,R_{\rm J}$  \citep{Diaz2014}. That object has an inferred age of 6 Gyr and a radius 15\% larger than expected from evolutionary models, comparable to \ztfb. \citet{Diaz2014} attributed the inflation of this object to inhibition of convection by strong magnetic fields \citep[e.g.][]{Mullan2001}. Such a scenario could plausibly also apply to \ztfb, particularly given that tidal synchronization is expected to maintain a dynamo. 

We consider tides or irradiation less likely to be responsible for the BD's inflation, since there exist BDs that are more strongly irradiated and experience stronger tides and are not inflated \citep{Littlefair2014, Parsons2017}. One such case is the white dwarf + BD binary SDSS J1205-0242, which contains a BD with mass $51\,M_{\rm J}$ orbiting a hot WD ($T_{\rm eff} =23680 \pm 430$\,K) in an orbital period of only 1.19 hours. The luminosity of the WD is $L_{\rm WD} = 0.14\,L_{\odot}$, 80 times brighter than the M dwarf in \ztfb. The orbital separation is 5\% smaller, leading to an incident flux per unit area $\approx 90$ times higher than the flux on the BD in \ztfb. The incident flux must have been even higher in the past, when the 50 Myr-old WD was younger and hotter. Yet, the BD in SDSS J1205-0242 is not inflated: with a radius of only $0.82\pm 0.03\,R_{\rm J}$, it is one of the smallest known BDs. Some of the observed BDs orbiting solar-type stars with $P_{\rm orb}  = 1-3$\,d also experience a higher incident flux that the BD in \ztfbd on account of their much more luminous host stars \citep[e.g.][]{Sebastian2022}. Given that the amount of radius inflation in other massive BDs is also not strongly correlated with how much they are irradiated, we conclude that it is more likely to be a result of processes internal to the BD.

About a dozen BDs are known that are in short-period detached binaries with white dwarf or hot subdwarf companions \citep[e.g.][]{Burleigh2006, Geier2011, Steele2013, Parsons2017, Casewell2018, Casewell2020b, Casewell2020c}, and several of these have orbital periods of 1-2 hours. The critical difference between these systems and \ztfbd is that they are post-common envelope systems, meaning that the BD likely formed in wide orbit ($\sim $au) and only ended up at short periods after a common envelope event. There is no need to appeal to magnetic braking to explain the tight orbits of BDs in post-common envelope binaries. \ztfbd has the shortest confirmed orbital period among all non-post common envelope binaries \citep[whether they contain main-squence stars or BDs; e.g.][]{Nefs2012, Soszynski2015}. 

\begin{figure*}
    \centering
    \includegraphics[width=\textwidth]{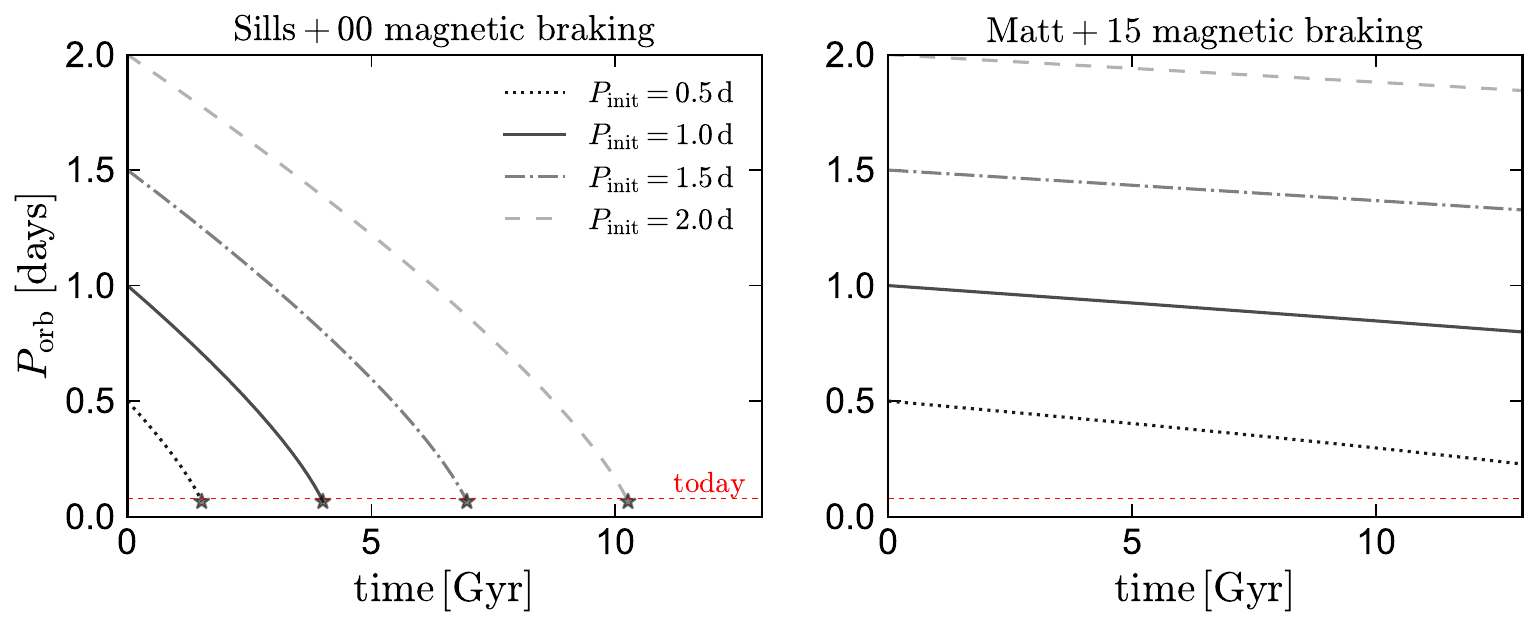}
    \caption{Predicted orbital period evolution of \ztfbd under two different magnetic braking laws.  In both panels, different tracks show four different initial periods, dashed red line shows the period of \ztfb, and grey stars show the onset of mass transfer. With the \citet{Sills2000} prescription (left panel), magnetic braking is fairly efficient and can explain the current orbit for any initial period below 2 days. \citet[][right panel]{Matt2015} predicts significantly weaker magnetic braking and cannot significantly shrink the orbit in a Hubble time.   }
    \label{fig:MB}
\end{figure*}

\subsection{Formation history}
\label{sec:formation_history}
The two components of \ztfbd both currently fit inside their Roche lobes, but both must have been significantly larger in youth. For example, at an age of 1 Myr, the M dwarf is predicted to have had a radius of $1.2\,R_{\odot}$ \citep[e.g.][]{Baraffe2015}; at this age, it would have overflowed its Roche lobe for any $P_{\rm orb}< 1.16\,$days. At an age of 10 Myr, the same figures are $0.49\,R_{\odot}$ and 0.31 days. The simplest explanation is thus that the binary formed with a wider orbit and subsequently shrunk via magnetic braking. Dynamical interactions with other bodies in the system's birth environment and/or an unseen tertiary component likely also played an important role in the binary's formation \citep{Tokovinin2006}, but such interactions become inefficient once tides dominate apsidal precession \citep[e.g.][]{Holman1997, Blaes2002, Fabrycky2007}. In the absence of magnetic braking or other sources of angular momentum loss, we thus do not expect binary orbits to shrink below 1-2 days \citep[e.g.][]{Stepien2006, Hwang2020}. Gravitational waves are too weak to explain the observed orbit: assuming an age of 10 Gyr, the initial period would have been only 0.14 days if they provided the only angular momentum loss \citep{Peters1964}. This suggests that magnetic braking remains efficient in at least some fully-convective stars.

To assess how the orbit may have shrunk, we calculated its expected evolution under two different magnetic braking prescriptions. The first is taken from \citet{Sills2000}, as motivated by \citet{Kawaler1988} and \citet{Chaboyer1995}, and the second from \citet{Matt2015}, based on the rotation period distributions of stars observed by {\it Kepler}. Both prescriptions place \ztfbd in the saturated regime, with $\dot{J}\propto P_{\rm orb}^{-1}$, but the calibration from \citet{Matt2015} declines more strongly at low masses and thus predicts weaker braking. 

Figure~\ref{fig:MB} shows the predicted orbital evolution of \ztfbd under both magnetic braking laws for four different initial periods. As described by \citet{El-Badry2022}, we assume both components are tidally synchronized and that each component removes angular momentum from the orbit following the same (mass- and radius-dependent) braking law. Tracks terminate at an orbital period of 1.56 hours, when the M dwarf will overflow its Roche lobe. 

The \citet{Sills2000} magnetic braking prescription (left panel) predicts relatively efficient angular momentum loss, such that any initial period below 2 days could reach the observed orbit within 10 Gyr. In this prescription, magnetic braking dominates over gravitational waves at all periods, such that the inspiral time until Roche lobe overflow is predicted to be only 30 Myr, compared to the predicted 1.3 Gyr due to gravitational waves alone. 

In contrast, the \citet{Matt2015} prescription (right panel) predicts magnetic braking to be rather weak, resulting in only small changes to the orbital period over a Hubble time. In this prescription the magnetic braking torque scales as $\dot{J}_{\rm MB}\propto R^{3.1}M^{0.5}$, resulting in very weak braking for low-mass stars and BDs. With such weak braking it is impossible to explain the current orbit of \ztfbd in the context of isolated binary evolution.

Given our conclusion that the orbit of \ztfbd likely decayed due to magnetic braking, we can consider whether their exist plausible progenitor systems with wider orbits. The recently discovered binary LP 413-53AB -- which was until now the shortest-period known binary containing a late M dwarf and a likely BD -- is a prime candidate. That system has  an orbital period of 17 hours and components similar to \ztfbd \citep{Hsu2023}. Its orbit will likely continue to shrink via magnetic braking, which in the \citet{Sills2000} prescription will bring it to a period of 1.9 hours within a few Gyr.

\subsection{Future evolution}
Because the BD is somewhat more dense than the M dwarf, it is the M dwarf that will first overflow its Roche lobe. Given the M dwarf's current radius, this will occur at an orbital period of 1.56 hours. The orbit is by now tight enough that gravitational wave-driven inspiral is important for the binary's future evolution, despite the low masses of the two components. If the orbit decays {\it only} due to gravitational waves, the M dwarf will overflow its Roche lobe in 1.3 Gyr. This represents a lower limit on the system's remaining lifetime as a detached binary, because magnetic braking likely contributes significant additional angular momentum loss. 

When the M dwarf overflows its Roche lobe, the closest approach of the accretion stream to the BD is predicted to be $r_{\rm min}\approx 0.15\,R_{\rm J}$ \citep{Lubow1975}. Because this is smaller than the radius of the BD, mass transfer is predicted to occur through direct impact accretion rather than through a disk. This will lead to the formation of a hot spot on the BD, most likely leading to significant changes in the optical light curve. 

The fate of \ztfbd following the onset of mass transfer is somewhat uncertain. Given that $M_\star > M_{\rm BD}$, mass transfer will at first proceed on the M dwarf's thermal timescale, which is of order 1 Gyr. Since the BD's thermal timescale is significantly longer than this, it may expand as a result of accretion, potentially resulting in the formation of a common-envelope and a long-lived contact binary. If the system survives, it would be a good candidate to form an ultramassive BD more massive than the hydrogen burning limit \citep[e.g.][]{Salpeter1992, Forbes2019}.

There very likely exist analogs of \ztfbd in which {\it both} components are BDs. Such systems would be very faint in the optical while detached, but might appear as high-amplitude variables in the optical due to their rotating hot spots once mass transfer begins. Indeed, several such sources have recently been discovered in ZTF data. Their properties and possible evolutionary link to detached binaries like \ztfbd will be explored in future work.

\subsection{Prospects for finding more short-period BDs}
At a distance of only $\approx 140$\,pc, \ztfbd is closer than 34 of the 39 other known transiting BDs shown in Figure~\ref{fig:pop}. This suggests that BDs in short-period orbits are not particularly rare. However, \ztfbd is almost 2 magnitudes fainter in the $G$ band than any of those systems. This reflects the fact that the host star has a high density and low mass -- as it must in order to fit inside a tight orbit. Most of the other known transiting BDs were discovered by surveys like {\it TESS}  {\it Kepler}, {\it CoRoT}, and WASP, which have limited sensitivity to faint, low-mass stars. Deeper forthcoming surveys such as WINTER \citep{Lourie2020}, Rubin \citep{Ivezic2019}, and {\it Roman} \citep{Spergel2015} thus have bright prospects for detecting more BDs in short-period orbits. 

\section{Summary and conclusion}
\label{sec:conclusion}
We have discovered a new detached binary containing a low-mass star ($M_\star = 0.134 \pm 0.004\,M_{\odot}$; spectral type M6) and a high-mass BD ($M_{\rm BD} = 80.1 \pm 1.6\,M_{\rm J}$; spectral type L5). Joint fitting of the system's spectral energy distribution (Figures~\ref{fig:summary} and~\ref{fig:sptyep})  multi-band light curves (Figure~\ref{fig:lcs}), and radial velocities (Figure~\ref{fig:river}) allow us to place tight constraint on the physical parameters of both components (Figures~\ref{fig:fitting_results} and~\ref{fig:corner}). We find that the BD is mildly inflated and has a mass just below the hydrogen burning limit (Figure~\ref{fig:evols}). 

With an orbital period of only 1.90 hours, the system is much more compact than other known transiting brown dwarfs (Figure~\ref{fig:pop}). Both components must have been significantly larger when they were young than they are today, implying that the orbit has shrunk significantly by magnetic braking. This strongly suggests that magnetic braking remains efficient below the fully convective boundary in at least some stars (Figure~\ref{fig:MB}), contrary to the common assumption in many binary evolution models. 

The system's orbit will continue to decay in the future, and the M dwarf will overflow its Roche lobe in 1.3 Gyr if gravitational radiation is the only significant angular momentum loss mechanism. In the likely event that magnetic braking leads to additional angular momentum loss, mass transfer is expected to begin within a few tens of Myr.

\section*{acknowledgments}
We thank the anonymous referee for a constructive report, and Jackie Faherty, Mercedes López-Morales, and David Charbonneau for useful discussions. 

Based on observations obtained with the Samuel Oschin Telescope 48-inch and the 60-inch Telescope at the Palomar
Observatory as part of the Zwicky Transient Facility project. ZTF is supported by the National Science Foundation under Grant
No. AST-2034437 and a collaboration including Caltech, IPAC, the Weizmann Institute for Science, the Oskar Klein Center at
Stockholm University, the University of Maryland, Deutsches Elektronen-Synchrotron and Humboldt University, the TANGO
Consortium of Taiwan, the University of Wisconsin at Milwaukee, Trinity College Dublin, Lawrence Livermore National
Laboratories, and IN2P3, France. Operations are conducted by COO, IPAC, and UW.

The data presented herein were obtained at the W. M. Keck Observatory, which is operated as a scientific partnership among the California Institute of Technology, the University of California and the National Aeronautics and Space Administration. The Observatory was made possible by the generous financial support of the W. M. Keck Foundation. The authors wish to recognize and acknowledge the very significant cultural role and reverence that the summit of Maunakea has always had within the indigenous Hawaiian community.  We are most fortunate to have the opportunity to conduct observations from this mountain.

\newpage

\newpage

\bibliographystyle{mnras}

\begin{thebibliography}{}
\makeatletter
\relax
\def\mn@urlcharsother{\let\do\@makeother \do\$\do\&\do\#\do\^\do\_\do\%\do\~}
\def\mn@doi{\begingroup\mn@urlcharsother \@ifnextchar [ {\mn@doi@}
  {\mn@doi@[]}}
\def\mn@doi@[#1]#2{\def\@tempa{#1}\ifx\@tempa\@empty \href
  {http://dx.doi.org/#2} {doi:#2}\else \href {http://dx.doi.org/#2} {#1}\fi
  \endgroup}
\def\mn@eprint#1#2{\mn@eprint@#1:#2::\@nil}
\def\mn@eprint@arXiv#1{\href {http://arxiv.org/abs/#1} {{\tt arXiv:#1}}}
\def\mn@eprint@dblp#1{\href {http://dblp.uni-trier.de/rec/bibtex/#1.xml}
  {dblp:#1}}
\def\mn@eprint@#1:#2:#3:#4\@nil{\def\@tempa {#1}\def\@tempb {#2}\def\@tempc
  {#3}\ifx \@tempc \@empty \let \@tempc \@tempb \let \@tempb \@tempa \fi \ifx
  \@tempb \@empty \def\@tempb {arXiv}\fi \@ifundefined
  {mn@eprint@\@tempb}{\@tempb:\@tempc}{\expandafter \expandafter \csname
  mn@eprint@\@tempb\endcsname \expandafter{\@tempc}}}

\bibitem[\protect\citeauthoryear{{Acton} et~al.,}{{Acton}
  et~al.}{2021}]{Acton2021}
{Acton} J.~S.,  et~al., 2021, \mn@doi [\mnras] {10.1093/mnras/stab1459}, \href
  {https://ui.adsabs.harvard.edu/abs/2021MNRAS.505.2741A} {505, 2741}

\bibitem[\protect\citeauthoryear{{Aganze} et~al.,}{{Aganze}
  et~al.}{2022}]{Aganze2022}
{Aganze} C.,  et~al., 2022, \mn@doi [\apj] {10.3847/1538-4357/ac35ea}, \href
  {https://ui.adsabs.harvard.edu/abs/2022ApJ...924..114A} {924, 114}

\bibitem[\protect\citeauthoryear{{Allard}, {Homeier}  \& {Freytag}}{{Allard}
  et~al.}{2011}]{Allard2011}
{Allard} F.,  {Homeier} D.,   {Freytag} B.,  2011, in {Johns-Krull} C.,
  {Browning} M.~K.,   {West} A.~A.,  eds,  Astronomical Society of the Pacific
  Conference Series Vol. 448, 16th Cambridge Workshop on Cool Stars, Stellar
  Systems, and the Sun. p.~91 (\mn@eprint {arXiv} {1011.5405}),
  \mn@doi{10.48550/arXiv.1011.5405}

\bibitem[\protect\citeauthoryear{{Applegate}}{{Applegate}}{1992}]{Applegate1992}
{Applegate} J.~H.,  1992, \mn@doi [\apj] {10.1086/170967}, \href
  {https://ui.adsabs.harvard.edu/abs/1992ApJ...385..621A} {385, 621}

\bibitem[\protect\citeauthoryear{{Baglin} et~al.,}{{Baglin}
  et~al.}{2006}]{Baglin2006}
{Baglin} A.,  et~al., 2006, in 36th COSPAR Scientific Assembly. p.~3749

\bibitem[\protect\citeauthoryear{{Baraffe} \& {Chabrier}}{{Baraffe} \&
  {Chabrier}}{1996}]{Baraffe1996}
{Baraffe} I.,  {Chabrier} G.,  1996, \mn@doi [\apjl] {10.1086/309988}, \href
  {https://ui.adsabs.harvard.edu/abs/1996ApJ...461L..51B} {461, L51}

\bibitem[\protect\citeauthoryear{{Baraffe}, {Chabrier}, {Allard}  \&
  {Hauschildt}}{{Baraffe} et~al.}{2002}]{Baraffe2002}
{Baraffe} I.,  {Chabrier} G.,  {Allard} F.,   {Hauschildt} P.~H.,  2002,
  \mn@doi [\aap] {10.1051/0004-6361:20011638}, \href
  {https://ui.adsabs.harvard.edu/abs/2002A&A...382..563B} {382, 563}

\bibitem[\protect\citeauthoryear{{Baraffe}, {Homeier}, {Allard}  \&
  {Chabrier}}{{Baraffe} et~al.}{2015}]{Baraffe2015}
{Baraffe} I.,  {Homeier} D.,  {Allard} F.,   {Chabrier} G.,  2015, \mn@doi
  [\aap] {10.1051/0004-6361/201425481}, \href
  {https://ui.adsabs.harvard.edu/abs/2015A&A...577A..42B} {577, A42}

\bibitem[\protect\citeauthoryear{{Bayless} \& {Orosz}}{{Bayless} \&
  {Orosz}}{2006}]{Bayless2006}
{Bayless} A.~J.,  {Orosz} J.~A.,  2006, \mn@doi [\apj] {10.1086/507981}, \href
  {https://ui.adsabs.harvard.edu/abs/2006ApJ...651.1155B} {651, 1155}

\bibitem[\protect\citeauthoryear{{Bellm} et~al.,}{{Bellm}
  et~al.}{2019}]{Bellm2019}
{Bellm} E.~C.,  et~al., 2019, \mn@doi [\pasp] {10.1088/1538-3873/aaecbe}, \href
  {https://ui.adsabs.harvard.edu/abs/2019PASP..131a8002B} {131, 018002}

\bibitem[\protect\citeauthoryear{{Blaes}, {Lee}  \& {Socrates}}{{Blaes}
  et~al.}{2002}]{Blaes2002}
{Blaes} O.,  {Lee} M.~H.,   {Socrates} A.,  2002, \mn@doi [\apj]
  {10.1086/342655}, \href
  {https://ui.adsabs.harvard.edu/abs/2002ApJ...578..775B} {578, 775}

\bibitem[\protect\citeauthoryear{{Borucki} et~al.,}{{Borucki}
  et~al.}{2010}]{Borucki2010}
{Borucki} W.~J.,  et~al., 2010, \mn@doi [Science] {10.1126/science.1185402},
  \href {https://ui.adsabs.harvard.edu/abs/2010Sci...327..977B} {327, 977}

\bibitem[\protect\citeauthoryear{{Bouchy}, {Deleuil}, {Guillot}, {Aigrain},
  {Carone}  \& {Cochran}}{{Bouchy} et~al.}{2010}]{Bouchy2010}
{Bouchy} F.,  {Deleuil} M.,  {Guillot} T.,  {Aigrain} S.,  {Carone} L.,
  {Cochran} W.~D.,  2010, \mn@doi [arXiv e-prints] {10.48550/arXiv.1010.0179},
  \href {https://ui.adsabs.harvard.edu/abs/2010arXiv1010.0179B} {p.
  arXiv:1010.0179}

\bibitem[\protect\citeauthoryear{{Bovy}}{{Bovy}}{2015}]{Bovy2015}
{Bovy} J.,  2015, \mn@doi [\apjs] {10.1088/0067-0049/216/2/29}, \href
  {https://ui.adsabs.harvard.edu/abs/2015ApJS..216...29B} {216, 29}

\bibitem[\protect\citeauthoryear{{Burleigh}, {Hogan}, {Dobbie}, {Napiwotzki}
  \& {Maxted}}{{Burleigh} et~al.}{2006}]{Burleigh2006}
{Burleigh} M.~R.,  {Hogan} E.,  {Dobbie} P.~D.,  {Napiwotzki} R.,   {Maxted}
  P.~F.~L.,  2006, \mn@doi [\mnras] {10.1111/j.1745-3933.2006.00242.x}, \href
  {https://ui.adsabs.harvard.edu/abs/2006MNRAS.373L..55B} {373, L55}

\bibitem[\protect\citeauthoryear{{Burrows}, {Hubbard}, {Saumon}  \&
  {Lunine}}{{Burrows} et~al.}{1993}]{Burrows1993}
{Burrows} A.,  {Hubbard} W.~B.,  {Saumon} D.,   {Lunine} J.~I.,  1993, \mn@doi
  [\apj] {10.1086/172427}, \href
  {https://ui.adsabs.harvard.edu/abs/1993ApJ...406..158B} {406, 158}

\bibitem[\protect\citeauthoryear{{Carmichael}}{{Carmichael}}{2023}]{Carmichael2023}
{Carmichael} T.~W.,  2023, \mn@doi [\mnras] {10.1093/mnras/stac3720}, \href
  {https://ui.adsabs.harvard.edu/abs/2023MNRAS.519.5177C} {519, 5177}

\bibitem[\protect\citeauthoryear{{Carmichael} et~al.,}{{Carmichael}
  et~al.}{2020}]{Carmichael2020}
{Carmichael} T.~W.,  et~al., 2020, \mn@doi [\aj] {10.3847/1538-3881/ab9b84},
  \href {https://ui.adsabs.harvard.edu/abs/2020AJ....160...53C} {160, 53}

\bibitem[\protect\citeauthoryear{{Casewell} et~al.,}{{Casewell}
  et~al.}{2018}]{Casewell2018}
{Casewell} S.~L.,  et~al., 2018, \mn@doi [\mnras] {10.1093/mnras/sty245}, \href
  {https://ui.adsabs.harvard.edu/abs/2018MNRAS.476.1405C} {476, 1405}

\bibitem[\protect\citeauthoryear{{Casewell} et~al.,}{{Casewell}
  et~al.}{2020a}]{Casewell2020b}
{Casewell} S.~L.,  et~al., 2020a, \mn@doi [\mnras] {10.1093/mnras/staa1608},
  \href {https://ui.adsabs.harvard.edu/abs/2020MNRAS.497.3571C} {497, 3571}

\bibitem[\protect\citeauthoryear{{Casewell} et~al.,}{{Casewell}
  et~al.}{2020b}]{Casewell2020c}
{Casewell} S.~L.,  et~al., 2020b, \mn@doi [\mnras] {10.1093/mnras/staa1608},
  \href {https://ui.adsabs.harvard.edu/abs/2020MNRAS.497.3571C} {497, 3571}

\bibitem[\protect\citeauthoryear{{Casewell}, {Debes}, {Braker}, {Cushing},
  {Mace}, {Marley}  \& {Kirkpatrick}}{{Casewell} et~al.}{2020c}]{Casewell2020}
{Casewell} S.~L.,  {Debes} J.,  {Braker} I.~P.,  {Cushing} M.~C.,  {Mace} G.,
  {Marley} M.~S.,   {Kirkpatrick} J.~D.,  2020c, \mn@doi [\mnras]
  {10.1093/mnras/staa3184}, \href
  {https://ui.adsabs.harvard.edu/abs/2020MNRAS.499.5318C} {499, 5318}

\bibitem[\protect\citeauthoryear{{Chaboyer}, {Demarque}  \&
  {Pinsonneault}}{{Chaboyer} et~al.}{1995}]{Chaboyer1995}
{Chaboyer} B.,  {Demarque} P.,   {Pinsonneault} M.~H.,  1995, \mn@doi [\apj]
  {10.1086/175409}, \href
  {https://ui.adsabs.harvard.edu/abs/1995ApJ...441..876C} {441, 876}

\bibitem[\protect\citeauthoryear{{Chabrier} \& {Baraffe}}{{Chabrier} \&
  {Baraffe}}{1997}]{Chabrier1997}
{Chabrier} G.,  {Baraffe} I.,  1997, \mn@doi [\aap]
  {10.48550/arXiv.astro-ph/9704118}, \href
  {https://ui.adsabs.harvard.edu/abs/1997A&A...327.1039C} {327, 1039}

\bibitem[\protect\citeauthoryear{{Chabrier}, {Baraffe}, {Phillips}  \&
  {Debras}}{{Chabrier} et~al.}{2023}]{Chabrier2023}
{Chabrier} G.,  {Baraffe} I.,  {Phillips} M.,   {Debras} F.,  2023, \mn@doi
  [\aap] {10.1051/0004-6361/202243832}, \href
  {https://ui.adsabs.harvard.edu/abs/2023A&A...671A.119C} {671, A119}

\bibitem[\protect\citeauthoryear{{Claret}, {Hauschildt}  \& {Witte}}{{Claret}
  et~al.}{2012}]{Claret2012}
{Claret} A.,  {Hauschildt} P.~H.,   {Witte} S.,  2012, \mn@doi [\aap]
  {10.1051/0004-6361/201219849}, \href
  {https://ui.adsabs.harvard.edu/abs/2012A&A...546A..14C} {546, A14}

\bibitem[\protect\citeauthoryear{{Covey} et~al.,}{{Covey}
  et~al.}{2007}]{Covey2007}
{Covey} K.~R.,  et~al., 2007, \mn@doi [\aj] {10.1086/522052}, \href
  {https://ui.adsabs.harvard.edu/abs/2007AJ....134.2398C} {134, 2398}

\bibitem[\protect\citeauthoryear{{Cruz}, {Diaz}, {Birkby}, {Barrado},
  {Sip{\"o}cz}  \& {Hodgkin}}{{Cruz} et~al.}{2018}]{Cruz2018}
{Cruz} P.,  {Diaz} M.,  {Birkby} J.,  {Barrado} D.,  {Sip{\"o}cz} B.,
  {Hodgkin} S.,  2018, \mn@doi [\mnras] {10.1093/mnras/sty541}, \href
  {https://ui.adsabs.harvard.edu/abs/2018MNRAS.476.5253C} {476, 5253}

\bibitem[\protect\citeauthoryear{{D{\'\i}az} et~al.,}{{D{\'\i}az}
  et~al.}{2014}]{Diaz2014}
{D{\'\i}az} R.~F.,  et~al., 2014, \mn@doi [\aap] {10.1051/0004-6361/201424406},
  \href {https://ui.adsabs.harvard.edu/abs/2014A&A...572A.109D} {572, A109}

\bibitem[\protect\citeauthoryear{{Dieterich}, {Henry}, {Jao}, {Winters},
  {Hosey}, {Riedel}  \& {Subasavage}}{{Dieterich} et~al.}{2014}]{Dieterich2014}
{Dieterich} S.~B.,  {Henry} T.~J.,  {Jao} W.-C.,  {Winters} J.~G.,  {Hosey}
  A.~D.,  {Riedel} A.~R.,   {Subasavage} J.~P.,  2014, \mn@doi [\aj]
  {10.1088/0004-6256/147/5/94}, \href
  {https://ui.adsabs.harvard.edu/abs/2014AJ....147...94D} {147, 94}

\bibitem[\protect\citeauthoryear{{Dupuy} \& {Liu}}{{Dupuy} \&
  {Liu}}{2017}]{Dupuy2017}
{Dupuy} T.~J.,  {Liu} M.~C.,  2017, \mn@doi [\apjs] {10.3847/1538-4365/aa5e4c},
  \href {https://ui.adsabs.harvard.edu/abs/2017ApJS..231...15D} {231, 15}

\bibitem[\protect\citeauthoryear{{El-Badry}, {Rix}  \& {Heintz}}{{El-Badry}
  et~al.}{2021}]{El-Badry2021}
{El-Badry} K.,  {Rix} H.-W.,   {Heintz} T.~M.,  2021, \mn@doi [\mnras]
  {10.1093/mnras/stab323}, \href
  {https://ui.adsabs.harvard.edu/abs/2021MNRAS.506.2269E} {506, 2269}

\bibitem[\protect\citeauthoryear{{El-Badry}, {Conroy}, {Fuller}, {Kiman}, {van
  Roestel}, {Rodriguez}  \& {Burdge}}{{El-Badry} et~al.}{2022}]{El-Badry2022}
{El-Badry} K.,  {Conroy} C.,  {Fuller} J.,  {Kiman} R.,  {van Roestel} J.,
  {Rodriguez} A.~C.,   {Burdge} K.~B.,  2022, \mn@doi [\mnras]
  {10.1093/mnras/stac2945}, \href
  {https://ui.adsabs.harvard.edu/abs/2022MNRAS.517.4916E} {517, 4916}

\bibitem[\protect\citeauthoryear{{Fabrycky} \& {Tremaine}}{{Fabrycky} \&
  {Tremaine}}{2007}]{Fabrycky2007}
{Fabrycky} D.,  {Tremaine} S.,  2007, \mn@doi [\apj] {10.1086/521702}, \href
  {https://ui.adsabs.harvard.edu/abs/2007ApJ...669.1298F} {669, 1298}

\bibitem[\protect\citeauthoryear{{Forbes} \& {Loeb}}{{Forbes} \&
  {Loeb}}{2019}]{Forbes2019}
{Forbes} J.~C.,  {Loeb} A.,  2019, \mn@doi [\apj] {10.3847/1538-4357/aafac8},
  \href {https://ui.adsabs.harvard.edu/abs/2019ApJ...871..227F} {871, 227}

\bibitem[\protect\citeauthoryear{{Foreman-Mackey}, {Hogg}, {Lang}  \&
  {Goodman}}{{Foreman-Mackey} et~al.}{2013}]{Foreman-Mackey2013}
{Foreman-Mackey} D.,  {Hogg} D.~W.,  {Lang} D.,   {Goodman} J.,  2013, \mn@doi
  [\pasp] {10.1086/670067}, \href
  {https://ui.adsabs.harvard.edu/abs/2013PASP..125..306F} {125, 306}

\bibitem[\protect\citeauthoryear{{Gaia Collaboration} et~al.,}{{Gaia
  Collaboration} et~al.}{2021a}]{GaiaCollaboration2021}
{Gaia Collaboration} et~al., 2021a, \mn@doi [\aap]
  {10.1051/0004-6361/202039657}, \href
  {https://ui.adsabs.harvard.edu/abs/2021A&A...649A...1G} {649, A1}

\bibitem[\protect\citeauthoryear{{Gaia Collaboration} et~al.,}{{Gaia
  Collaboration} et~al.}{2021b}]{Smart2021}
{Gaia Collaboration} et~al., 2021b, \mn@doi [\aap]
  {10.1051/0004-6361/202039498}, \href
  {https://ui.adsabs.harvard.edu/abs/2021A&A...649A...6G} {649, A6}

\bibitem[\protect\citeauthoryear{{Gaia Collaboration} et~al.,}{{Gaia
  Collaboration} et~al.}{2022}]{GaiaCollaboration2022}
{Gaia Collaboration} et~al., 2022, \mn@doi [arXiv e-prints]
  {10.48550/arXiv.2208.00211}, \href
  {https://ui.adsabs.harvard.edu/abs/2022arXiv220800211G} {p. arXiv:2208.00211}

\bibitem[\protect\citeauthoryear{{Garraffo}, {Drake}  \& {Cohen}}{{Garraffo}
  et~al.}{2015}]{Garraffo2015}
{Garraffo} C.,  {Drake} J.~J.,   {Cohen} O.,  2015, \mn@doi [\apjl]
  {10.1088/2041-8205/807/1/L6}, \href
  {https://ui.adsabs.harvard.edu/abs/2015ApJ...807L...6G} {807, L6}

\bibitem[\protect\citeauthoryear{{Geier} et~al.,}{{Geier}
  et~al.}{2011}]{Geier2011}
{Geier} S.,  et~al., 2011, \mn@doi [\apjl] {10.1088/2041-8205/731/2/L22}, \href
  {https://ui.adsabs.harvard.edu/abs/2011ApJ...731L..22G} {731, L22}

\bibitem[\protect\citeauthoryear{{Gelino}, {Kirkpatrick}  \&
  {Burgasser}}{{Gelino} et~al.}{2009}]{Gelino2009}
{Gelino} C.~R.,  {Kirkpatrick} J.~D.,   {Burgasser} A.~J.,  2009, in {Stempels}
  E.,  ed.,  American Institute of Physics Conference Series Vol. 1094, 15th
  Cambridge Workshop on Cool Stars, Stellar Systems, and the Sun. pp 924--927,
  \mn@doi{10.1063/1.3099269}

\bibitem[\protect\citeauthoryear{{Grether} \& {Lineweaver}}{{Grether} \&
  {Lineweaver}}{2006}]{Grether2006}
{Grether} D.,  {Lineweaver} C.~H.,  2006, \mn@doi [\apj] {10.1086/500161},
  \href {https://ui.adsabs.harvard.edu/abs/2006ApJ...640.1051G} {640, 1051}

\bibitem[\protect\citeauthoryear{{Grieves} et~al.,}{{Grieves}
  et~al.}{2021}]{Grieves2021}
{Grieves} N.,  et~al., 2021, \mn@doi [\aap] {10.1051/0004-6361/202141145},
  \href {https://ui.adsabs.harvard.edu/abs/2021A&A...652A.127G} {652, A127}

\bibitem[\protect\citeauthoryear{{Gullikson}, {Dodson-Robinson}  \&
  {Kraus}}{{Gullikson} et~al.}{2014}]{Gullikson2014}
{Gullikson} K.,  {Dodson-Robinson} S.,   {Kraus} A.,  2014, \mn@doi [\aj]
  {10.1088/0004-6256/148/3/53}, \href
  {https://ui.adsabs.harvard.edu/abs/2014AJ....148...53G} {148, 53}

\bibitem[\protect\citeauthoryear{{Harding} et~al.,}{{Harding}
  et~al.}{2016}]{chimera}
{Harding} L.~K.,  et~al., 2016, \mn@doi [\mnras] {10.1093/mnras/stw094}, \href
  {https://ui.adsabs.harvard.edu/abs/2016MNRAS.457.3036H} {457, 3036}

\bibitem[\protect\citeauthoryear{{Hod{\v{z}}i{\'c}} et~al.,}{{Hod{\v{z}}i{\'c}}
  et~al.}{2018}]{Hodzic2018}
{Hod{\v{z}}i{\'c}} V.,  et~al., 2018, \mn@doi [\mnras] {10.1093/mnras/sty2512},
  \href {https://ui.adsabs.harvard.edu/abs/2018MNRAS.481.5091H} {481, 5091}

\bibitem[\protect\citeauthoryear{{Holman}, {Touma}  \& {Tremaine}}{{Holman}
  et~al.}{1997}]{Holman1997}
{Holman} M.,  {Touma} J.,   {Tremaine} S.,  1997, \mn@doi [\nat]
  {10.1038/386254a0}, \href
  {https://ui.adsabs.harvard.edu/abs/1997Natur.386..254H} {386, 254}

\bibitem[\protect\citeauthoryear{{Hsu}, {Burgasser}  \& {Theissen}}{{Hsu}
  et~al.}{2023}]{Hsu2023}
{Hsu} C.-C.,  {Burgasser} A.~J.,   {Theissen} C.~A.,  2023, \mn@doi [\apjl]
  {10.3847/2041-8213/acba8c}, \href
  {https://ui.adsabs.harvard.edu/abs/2023ApJ...945L...6H} {945, L6}

\bibitem[\protect\citeauthoryear{{Hwang} \& {Zakamska}}{{Hwang} \&
  {Zakamska}}{2020}]{Hwang2020}
{Hwang} H.-C.,  {Zakamska} N.~L.,  2020, \mn@doi [\mnras]
  {10.1093/mnras/staa400}, \href
  {https://ui.adsabs.harvard.edu/abs/2020MNRAS.493.2271H} {493, 2271}

\bibitem[\protect\citeauthoryear{{Irwin} et~al.,}{{Irwin}
  et~al.}{2009}]{Irwin2009}
{Irwin} J.,  et~al., 2009, \mn@doi [\apj] {10.1088/0004-637X/701/2/1436}, \href
  {https://ui.adsabs.harvard.edu/abs/2009ApJ...701.1436I} {701, 1436}

\bibitem[\protect\citeauthoryear{{Ivezi{\'c}} et~al.,}{{Ivezi{\'c}}
  et~al.}{2019}]{Ivezic2019}
{Ivezi{\'c}} {\v{Z}}.,  et~al., 2019, \mn@doi [\apj]
  {10.3847/1538-4357/ab042c}, \href
  {https://ui.adsabs.harvard.edu/abs/2019ApJ...873..111I} {873, 111}

\bibitem[\protect\citeauthoryear{{Jackman} et~al.,}{{Jackman}
  et~al.}{2019}]{Jackman2019}
{Jackman} J. A.~G.,  et~al., 2019, \mn@doi [\mnras] {10.1093/mnras/stz2496},
  \href {https://ui.adsabs.harvard.edu/abs/2019MNRAS.489.5146J} {489, 5146}

\bibitem[\protect\citeauthoryear{{Jackson}, {Jeffries}, {Deliyannis}, {Sun}  \&
  {Douglas}}{{Jackson} et~al.}{2019}]{Jackson2019}
{Jackson} R.~J.,  {Jeffries} R.~D.,  {Deliyannis} C.~P.,  {Sun} Q.,   {Douglas}
  S.~T.,  2019, \mn@doi [\mnras] {10.1093/mnras/sty3184}, \href
  {https://ui.adsabs.harvard.edu/abs/2019MNRAS.483.1125J} {483, 1125}

\bibitem[\protect\citeauthoryear{{Jaehnig}, {Somers}  \& {Stassun}}{{Jaehnig}
  et~al.}{2019}]{Jaehnig2019}
{Jaehnig} K.,  {Somers} G.,   {Stassun} K.~G.,  2019, \mn@doi [\apj]
  {10.3847/1538-4357/ab21cf}, \href
  {https://ui.adsabs.harvard.edu/abs/2019ApJ...879...39J} {879, 39}

\bibitem[\protect\citeauthoryear{{Kaiser} et~al.,}{{Kaiser}
  et~al.}{2002}]{Kaiser2002}
{Kaiser} N.,  et~al., 2002, in {Tyson} J.~A.,  {Wolff} S.,  eds,  Society of
  Photo-Optical Instrumentation Engineers (SPIE) Conference Series Vol. 4836,
  Survey and Other Telescope Technologies and Discoveries. pp 154--164,
  \mn@doi{10.1117/12.457365}

\bibitem[\protect\citeauthoryear{{Kawaler}}{{Kawaler}}{1988}]{Kawaler1988}
{Kawaler} S.~D.,  1988, \mn@doi [\apj] {10.1086/166740}, \href
  {https://ui.adsabs.harvard.edu/abs/1988ApJ...333..236K} {333, 236}

\bibitem[\protect\citeauthoryear{{Kesseli}, {West}, {Veyette}, {Harrison},
  {Feldman}  \& {Bochanski}}{{Kesseli} et~al.}{2017}]{Kesseli2017}
{Kesseli} A.~Y.,  {West} A.~A.,  {Veyette} M.,  {Harrison} B.,  {Feldman} D.,
  {Bochanski} J.~J.,  2017, \mn@doi [\apjs] {10.3847/1538-4365/aa656d}, \href
  {https://ui.adsabs.harvard.edu/abs/2017ApJS..230...16K} {230, 16}

\bibitem[\protect\citeauthoryear{{Kesseli}, {Muirhead}, {Mann}  \&
  {Mace}}{{Kesseli} et~al.}{2018}]{Kesseli2018}
{Kesseli} A.~Y.,  {Muirhead} P.~S.,  {Mann} A.~W.,   {Mace} G.,  2018, \mn@doi
  [\aj] {10.3847/1538-3881/aabccb}, \href
  {https://ui.adsabs.harvard.edu/abs/2018AJ....155..225K} {155, 225}

\bibitem[\protect\citeauthoryear{{Kirkpatrick} et~al.,}{{Kirkpatrick}
  et~al.}{2021}]{Kirkpatrick2021}
{Kirkpatrick} J.~D.,  et~al., 2021, \mn@doi [\apjs] {10.3847/1538-4365/abd107},
  \href {https://ui.adsabs.harvard.edu/abs/2021ApJS..253....7K} {253, 7}

\bibitem[\protect\citeauthoryear{{Kov{\'a}cs}, {Zucker}  \&
  {Mazeh}}{{Kov{\'a}cs} et~al.}{2002}]{Kovacs2002}
{Kov{\'a}cs} G.,  {Zucker} S.,   {Mazeh} T.,  2002, \mn@doi [\aap]
  {10.1051/0004-6361:20020802}, \href
  {https://ui.adsabs.harvard.edu/abs/2002A&A...391..369K} {391, 369}

\bibitem[\protect\citeauthoryear{{Lanza}, {Rodono}  \& {Rosner}}{{Lanza}
  et~al.}{1998}]{Lanza1998}
{Lanza} A.~F.,  {Rodono} M.,   {Rosner} R.,  1998, \mn@doi [\mnras]
  {10.1046/j.1365-8711.1998.01446.x}, \href
  {https://ui.adsabs.harvard.edu/abs/1998MNRAS.296..893L} {296, 893}

\bibitem[\protect\citeauthoryear{{Littlefair} et~al.,}{{Littlefair}
  et~al.}{2014}]{Littlefair2014}
{Littlefair} S.~P.,  et~al., 2014, \mn@doi [\mnras] {10.1093/mnras/stu1895},
  \href {https://ui.adsabs.harvard.edu/abs/2014MNRAS.445.2106L} {445, 2106}

\bibitem[\protect\citeauthoryear{{L{\'o}pez-Morales} \&
  {Ribas}}{{L{\'o}pez-Morales} \& {Ribas}}{2005}]{Lopez-Morales2005}
{L{\'o}pez-Morales} M.,  {Ribas} I.,  2005, \mn@doi [\apj] {10.1086/432680},
  \href {https://ui.adsabs.harvard.edu/abs/2005ApJ...631.1120L} {631, 1120}

\bibitem[\protect\citeauthoryear{{Lourie} et~al.,}{{Lourie}
  et~al.}{2020}]{Lourie2020}
{Lourie} N.~P.,  et~al., 2020, in {Evans} C.~J.,  {Bryant} J.~J.,   {Motohara}
  K.,  eds,  Society of Photo-Optical Instrumentation Engineers (SPIE)
  Conference Series Vol. 11447, Ground-based and Airborne Instrumentation for
  Astronomy VIII. p. 114479K (\mn@eprint {arXiv} {2102.01109}),
  \mn@doi{10.1117/12.2561210}

\bibitem[\protect\citeauthoryear{{Lubow} \& {Shu}}{{Lubow} \&
  {Shu}}{1975}]{Lubow1975}
{Lubow} S.~H.,  {Shu} F.~H.,  1975, \mn@doi [\apj] {10.1086/153614}, \href
  {https://ui.adsabs.harvard.edu/abs/1975ApJ...198..383L} {198, 383}

\bibitem[\protect\citeauthoryear{{Ma} \& {Ge}}{{Ma} \& {Ge}}{2014}]{Ma2014}
{Ma} B.,  {Ge} J.,  2014, \mn@doi [\mnras] {10.1093/mnras/stu134}, \href
  {https://ui.adsabs.harvard.edu/abs/2014MNRAS.439.2781M} {439, 2781}

\bibitem[\protect\citeauthoryear{{Mace}}{{Mace}}{2014}]{Mace2014}
{Mace} G.~N.,  2014, PhD thesis, University of California, Los Angeles

\bibitem[\protect\citeauthoryear{{Mainzer} et~al.,}{{Mainzer}
  et~al.}{2014}]{Mainzer2014}
{Mainzer} A.,  et~al., 2014, \mn@doi [\apj] {10.1088/0004-637X/792/1/30}, \href
  {https://ui.adsabs.harvard.edu/abs/2014ApJ...792...30M} {792, 30}

\bibitem[\protect\citeauthoryear{{Mann}, {Brewer}, {Gaidos}, {L{\'e}pine}  \&
  {Hilton}}{{Mann} et~al.}{2013}]{Mann2013}
{Mann} A.~W.,  {Brewer} J.~M.,  {Gaidos} E.,  {L{\'e}pine} S.,   {Hilton}
  E.~J.,  2013, \mn@doi [\aj] {10.1088/0004-6256/145/2/52}, \href
  {https://ui.adsabs.harvard.edu/abs/2013AJ....145...52M} {145, 52}

\bibitem[\protect\citeauthoryear{{Mann} et~al.,}{{Mann}
  et~al.}{2019}]{Mann2019}
{Mann} A.~W.,  et~al., 2019, \mn@doi [\apj] {10.3847/1538-4357/aaf3bc}, \href
  {https://ui.adsabs.harvard.edu/abs/2019ApJ...871...63M} {871, 63}

\bibitem[\protect\citeauthoryear{{Marcy} \& {Butler}}{{Marcy} \&
  {Butler}}{2000}]{Marcy2000}
{Marcy} G.~W.,  {Butler} R.~P.,  2000, \mn@doi [\pasp] {10.1086/316516}, \href
  {https://ui.adsabs.harvard.edu/abs/2000PASP..112..137M} {112, 137}

\bibitem[\protect\citeauthoryear{{Marley} et~al.,}{{Marley}
  et~al.}{2021}]{Marley2021}
{Marley} M.~S.,  et~al., 2021, \mn@doi [\apj] {10.3847/1538-4357/ac141d}, \href
  {https://ui.adsabs.harvard.edu/abs/2021ApJ...920...85M} {920, 85}

\bibitem[\protect\citeauthoryear{{Matt}, {Brun}, {Baraffe}, {Bouvier}  \&
  {Chabrier}}{{Matt} et~al.}{2015}]{Matt2015}
{Matt} S.~P.,  {Brun} A.~S.,  {Baraffe} I.,  {Bouvier} J.,   {Chabrier} G.,
  2015, \mn@doi [\apjl] {10.1088/2041-8205/799/2/L23}, \href
  {https://ui.adsabs.harvard.edu/abs/2015ApJ...799L..23M} {799, L23}

\bibitem[\protect\citeauthoryear{{Maxted}}{{Maxted}}{2016}]{Maxted2016}
{Maxted} P.~F.~L.,  2016, \mn@doi [\aap] {10.1051/0004-6361/201628579}, \href
  {https://ui.adsabs.harvard.edu/abs/2016A&A...591A.111M} {591, A111}

\bibitem[\protect\citeauthoryear{{McLaughlin}}{{McLaughlin}}{1924}]{McLaughlin1924}
{McLaughlin} D.~B.,  1924, \mn@doi [\apj] {10.1086/142826}, \href
  {https://ui.adsabs.harvard.edu/abs/1924ApJ....60...22M} {60, 22}

\bibitem[\protect\citeauthoryear{{McMillan}}{{McMillan}}{2017}]{McMillan2017}
{McMillan} P.~J.,  2017, \mn@doi [\mnras] {10.1093/mnras/stw2759}, \href
  {https://ui.adsabs.harvard.edu/abs/2017MNRAS.465...76M} {465, 76}

\bibitem[\protect\citeauthoryear{{Mullan} \& {MacDonald}}{{Mullan} \&
  {MacDonald}}{2001}]{Mullan2001}
{Mullan} D.~J.,  {MacDonald} J.,  2001, \mn@doi [\apj] {10.1086/322336}, \href
  {https://ui.adsabs.harvard.edu/abs/2001ApJ...559..353M} {559, 353}

\bibitem[\protect\citeauthoryear{{Nefs} et~al.,}{{Nefs}
  et~al.}{2012}]{Nefs2012}
{Nefs} S.~V.,  et~al., 2012, \mn@doi [\mnras]
  {10.1111/j.1365-2966.2012.21338.x}, \href
  {https://ui.adsabs.harvard.edu/abs/2012MNRAS.425..950N} {425, 950}

\bibitem[\protect\citeauthoryear{{Newton}, {Irwin}, {Charbonneau}, {Berlind},
  {Calkins}  \& {Mink}}{{Newton} et~al.}{2017}]{Newton2017}
{Newton} E.~R.,  {Irwin} J.,  {Charbonneau} D.,  {Berlind} P.,  {Calkins}
  M.~L.,   {Mink} J.,  2017, \mn@doi [\apj] {10.3847/1538-4357/834/1/85}, \href
  {https://ui.adsabs.harvard.edu/abs/2017ApJ...834...85N} {834, 85}

\bibitem[\protect\citeauthoryear{{Palle} et~al.,}{{Palle}
  et~al.}{2021}]{Palle2021}
{Palle} E.,  et~al., 2021, \mn@doi [\aap] {10.1051/0004-6361/202039937}, \href
  {https://ui.adsabs.harvard.edu/abs/2021A&A...650A..55P} {650, A55}

\bibitem[\protect\citeauthoryear{{Parsons} et~al.,}{{Parsons}
  et~al.}{2017}]{Parsons2017}
{Parsons} S.~G.,  et~al., 2017, \mn@doi [\mnras] {10.1093/mnras/stx1610}, \href
  {https://ui.adsabs.harvard.edu/abs/2017MNRAS.471..976P} {471, 976}

\bibitem[\protect\citeauthoryear{{Parsons} et~al.,}{{Parsons}
  et~al.}{2018}]{Parsons2018}
{Parsons} S.~G.,  et~al., 2018, \mn@doi [\mnras] {10.1093/mnras/sty2345}, \href
  {https://ui.adsabs.harvard.edu/abs/2018MNRAS.481.1083P} {481, 1083}

\bibitem[\protect\citeauthoryear{{Parviainen} et~al.,}{{Parviainen}
  et~al.}{2020}]{Parviainen2020}
{Parviainen} H.,  et~al., 2020, \mn@doi [\aap] {10.1051/0004-6361/201935958},
  \href {https://ui.adsabs.harvard.edu/abs/2020A&A...633A..28P} {633, A28}

\bibitem[\protect\citeauthoryear{{Peters}}{{Peters}}{1964}]{Peters1964}
{Peters} P.~C.,  1964, \mn@doi [Physical Review] {10.1103/PhysRev.136.B1224},
  \href {https://ui.adsabs.harvard.edu/abs/1964PhRv..136.1224P} {136, 1224}

\bibitem[\protect\citeauthoryear{{Phillips} et~al.,}{{Phillips}
  et~al.}{2020}]{Phillips2020}
{Phillips} M.~W.,  et~al., 2020, \mn@doi [\aap] {10.1051/0004-6361/201937381},
  \href {https://ui.adsabs.harvard.edu/abs/2020A&A...637A..38P} {637, A38}

\bibitem[\protect\citeauthoryear{{Pollacco} et~al.,}{{Pollacco}
  et~al.}{2006}]{Pollacco2006}
{Pollacco} D.~L.,  et~al., 2006, \mn@doi [\pasp] {10.1086/508556}, \href
  {https://ui.adsabs.harvard.edu/abs/2006PASP..118.1407P} {118, 1407}

\bibitem[\protect\citeauthoryear{{Pr{\v{s}}a} \& {Zwitter}}{{Pr{\v{s}}a} \&
  {Zwitter}}{2005}]{Prsa2005}
{Pr{\v{s}}a} A.,  {Zwitter} T.,  2005, \mn@doi [\apj] {10.1086/430591}, \href
  {https://ui.adsabs.harvard.edu/abs/2005ApJ...628..426P} {628, 426}

\bibitem[\protect\citeauthoryear{{Raghavan} et~al.,}{{Raghavan}
  et~al.}{2010}]{Raghavan2010}
{Raghavan} D.,  et~al., 2010, \mn@doi [\apjs] {10.1088/0067-0049/190/1/1},
  \href {https://ui.adsabs.harvard.edu/abs/2010ApJS..190....1R} {190, 1}

\bibitem[\protect\citeauthoryear{{Rappaport}, {Verbunt}  \& {Joss}}{{Rappaport}
  et~al.}{1983}]{Rappaport1983}
{Rappaport} S.,  {Verbunt} F.,   {Joss} P.~C.,  1983, \mn@doi [\apj]
  {10.1086/161569}, \href
  {https://ui.adsabs.harvard.edu/abs/1983ApJ...275..713R} {275, 713}

\bibitem[\protect\citeauthoryear{{Rappaport}, {Vanderburg}, {Schwab}  \&
  {Nelson}}{{Rappaport} et~al.}{2021}]{Rappaport2021}
{Rappaport} S.,  {Vanderburg} A.,  {Schwab} J.,   {Nelson} L.,  2021, \mn@doi
  [\apj] {10.3847/1538-4357/abf7b0}, \href
  {https://ui.adsabs.harvard.edu/abs/2021ApJ...913..118R} {913, 118}

\bibitem[\protect\citeauthoryear{{Reiners} \& {Basri}}{{Reiners} \&
  {Basri}}{2008}]{Reiners2008}
{Reiners} A.,  {Basri} G.,  2008, \mn@doi [\apj] {10.1086/590073}, \href
  {https://ui.adsabs.harvard.edu/abs/2008ApJ...684.1390R} {684, 1390}

\bibitem[\protect\citeauthoryear{{Ricker} et~al.,}{{Ricker}
  et~al.}{2015}]{Ricker2015}
{Ricker} G.~R.,  et~al., 2015, \mn@doi [Journal of Astronomical Telescopes,
  Instruments, and Systems] {10.1117/1.JATIS.1.1.014003}, \href
  {https://ui.adsabs.harvard.edu/abs/2015JATIS...1a4003R} {1, 014003}

\bibitem[\protect\citeauthoryear{{Rossiter}}{{Rossiter}}{1924}]{Rossiter1924}
{Rossiter} R.~A.,  1924, \mn@doi [\apj] {10.1086/142825}, \href
  {https://ui.adsabs.harvard.edu/abs/1924ApJ....60...15R} {60, 15}

\bibitem[\protect\citeauthoryear{{Rothman} et~al.,}{{Rothman}
  et~al.}{2009}]{Rothman2009}
{Rothman} L.~S.,  et~al., 2009, \mn@doi [\jqsrt] {10.1016/j.jqsrt.2009.02.013},
  \href {https://ui.adsabs.harvard.edu/abs/2009JQSRT.110..533R} {110, 533}

\bibitem[\protect\citeauthoryear{{Sainsbury-Martinez}, {Casewell},
  {Lothringer}, {Phillips}  \& {Tremblin}}{{Sainsbury-Martinez}
  et~al.}{2021}]{Sainsbury-Martinez2021}
{Sainsbury-Martinez} F.,  {Casewell} S.~L.,  {Lothringer} J.~D.,  {Phillips}
  M.~W.,   {Tremblin} P.,  2021, \mn@doi [\aap] {10.1051/0004-6361/202141637},
  \href {https://ui.adsabs.harvard.edu/abs/2021A&A...656A.128S} {656, A128}

\bibitem[\protect\citeauthoryear{{Salpeter}}{{Salpeter}}{1992}]{Salpeter1992}
{Salpeter} E.~E.,  1992, \mn@doi [\apj] {10.1086/171502}, \href
  {https://ui.adsabs.harvard.edu/abs/1992ApJ...393..258S} {393, 258}

\bibitem[\protect\citeauthoryear{{Schatzman}}{{Schatzman}}{1962}]{Schatzman1962}
{Schatzman} E.,  1962, Annales d'Astrophysique, \href
  {https://ui.adsabs.harvard.edu/abs/1962AnAp...25...18S} {25, 18}

\bibitem[\protect\citeauthoryear{{Schreiber} et~al.,}{{Schreiber}
  et~al.}{2010}]{Schreiber2010}
{Schreiber} M.~R.,  et~al., 2010, \mn@doi [\aap] {10.1051/0004-6361/201013990},
  \href {https://ui.adsabs.harvard.edu/abs/2010A&A...513L...7S} {513, L7}

\bibitem[\protect\citeauthoryear{{Seabroke} \& {Gilmore}}{{Seabroke} \&
  {Gilmore}}{2007}]{Seabroke2007}
{Seabroke} G.~M.,  {Gilmore} G.,  2007, \mn@doi [\mnras]
  {10.1111/j.1365-2966.2007.12210.x}, \href
  {https://ui.adsabs.harvard.edu/abs/2007MNRAS.380.1348S} {380, 1348}

\bibitem[\protect\citeauthoryear{{Sebastian} et~al.,}{{Sebastian}
  et~al.}{2022}]{Sebastian2022}
{Sebastian} D.,  et~al., 2022, \mn@doi [\mnras] {10.1093/mnras/stac2131}, \href
  {https://ui.adsabs.harvard.edu/abs/2022MNRAS.516..636S} {516, 636}

\bibitem[\protect\citeauthoryear{{Sharma} et~al.,}{{Sharma}
  et~al.}{2014}]{Sharma2014}
{Sharma} S.,  et~al., 2014, \mn@doi [\apj] {10.1088/0004-637X/793/1/51}, \href
  {https://ui.adsabs.harvard.edu/abs/2014ApJ...793...51S} {793, 51}

\bibitem[\protect\citeauthoryear{{Sheinis}, {Bolte}, {Epps}, {Kibrick},
  {Miller}, {Radovan}, {Bigelow}  \& {Sutin}}{{Sheinis}
  et~al.}{2002}]{Sheinis2002}
{Sheinis} A.~I.,  {Bolte} M.,  {Epps} H.~W.,  {Kibrick} R.~I.,  {Miller} J.~S.,
   {Radovan} M.~V.,  {Bigelow} B.~C.,   {Sutin} B.~M.,  2002, \mn@doi [\pasp]
  {10.1086/341706}, \href
  {https://ui.adsabs.harvard.edu/abs/2002PASP..114..851S} {114, 851}

\bibitem[\protect\citeauthoryear{{Sills}, {Pinsonneault}  \&
  {Terndrup}}{{Sills} et~al.}{2000}]{Sills2000}
{Sills} A.,  {Pinsonneault} M.~H.,   {Terndrup} D.~M.,  2000, \mn@doi [\apj]
  {10.1086/308739}, \href
  {https://ui.adsabs.harvard.edu/abs/2000ApJ...534..335S} {534, 335}

\bibitem[\protect\citeauthoryear{{Skrutskie} et~al.,}{{Skrutskie}
  et~al.}{2006}]{Skrutskie2006}
{Skrutskie} M.~F.,  et~al., 2006, \mn@doi [\aj] {10.1086/498708}, \href
  {https://ui.adsabs.harvard.edu/abs/2006AJ....131.1163S} {131, 1163}

\bibitem[\protect\citeauthoryear{{Soszy{\'n}ski} et~al.,}{{Soszy{\'n}ski}
  et~al.}{2015}]{Soszynski2015}
{Soszy{\'n}ski} I.,  et~al., 2015, \mn@doi [\actaa]
  {10.48550/arXiv.1503.02080}, \href
  {https://ui.adsabs.harvard.edu/abs/2015AcA....65...39S} {65, 39}

\bibitem[\protect\citeauthoryear{{Spergel} et~al.,}{{Spergel}
  et~al.}{2015}]{Spergel2015}
{Spergel} D.,  et~al., 2015, \mn@doi [arXiv e-prints]
  {10.48550/arXiv.1503.03757}, \href
  {https://ui.adsabs.harvard.edu/abs/2015arXiv150303757S} {p. arXiv:1503.03757}

\bibitem[\protect\citeauthoryear{{Stamatellos} \& {Whitworth}}{{Stamatellos} \&
  {Whitworth}}{2009}]{Stamatellos2009}
{Stamatellos} D.,  {Whitworth} A.~P.,  2009, \mn@doi [\mnras]
  {10.1111/j.1365-2966.2008.14069.x}, \href
  {https://ui.adsabs.harvard.edu/abs/2009MNRAS.392..413S} {392, 413}

\bibitem[\protect\citeauthoryear{{Steele} et~al.,}{{Steele}
  et~al.}{2013}]{Steele2013}
{Steele} P.~R.,  et~al., 2013, \mn@doi [\mnras] {10.1093/mnras/sts620}, \href
  {https://ui.adsabs.harvard.edu/abs/2013MNRAS.429.3492S} {429, 3492}

\bibitem[\protect\citeauthoryear{{Stepien}}{{Stepien}}{2006}]{Stepien2006}
{Stepien} K.,  2006, \mn@doi [\actaa] {10.48550/arXiv.astro-ph/0701529}, \href
  {https://ui.adsabs.harvard.edu/abs/2006AcA....56..347S} {56, 347}

\bibitem[\protect\citeauthoryear{{Tokovinin}, {Thomas}, {Sterzik}  \&
  {Udry}}{{Tokovinin} et~al.}{2006}]{Tokovinin2006}
{Tokovinin} A.,  {Thomas} S.,  {Sterzik} M.,   {Udry} S.,  2006, \mn@doi [\aap]
  {10.1051/0004-6361:20054427}, \href
  {https://ui.adsabs.harvard.edu/abs/2006A&A...450..681T} {450, 681}

\bibitem[\protect\citeauthoryear{{Triaud} et~al.,}{{Triaud}
  et~al.}{2017}]{Triaud2017}
{Triaud} A. H.~M.~J.,  et~al., 2017, \mn@doi [\aap]
  {10.1051/0004-6361/201730993}, \href
  {https://ui.adsabs.harvard.edu/abs/2017A&A...608A.129T} {608, A129}

\bibitem[\protect\citeauthoryear{{Watson} \& {Marsh}}{{Watson} \&
  {Marsh}}{2010}]{Watson2010}
{Watson} C.~A.,  {Marsh} T.~R.,  2010, \mn@doi [\mnras]
  {10.1111/j.1365-2966.2010.16602.x}, \href
  {https://ui.adsabs.harvard.edu/abs/2010MNRAS.405.2037W} {405, 2037}

\bibitem[\protect\citeauthoryear{{Wright} et~al.,}{{Wright}
  et~al.}{2010}]{Wright2010}
{Wright} E.~L.,  et~al., 2010, \mn@doi [\aj] {10.1088/0004-6256/140/6/1868},
  \href {https://ui.adsabs.harvard.edu/abs/2010AJ....140.1868W} {140, 1868}

\bibitem[\protect\citeauthoryear{{Yu} \& {Liu}}{{Yu} \& {Liu}}{2018}]{Yu2018}
{Yu} J.,  {Liu} C.,  2018, \mn@doi [\mnras] {10.1093/mnras/stx3204}, \href
  {https://ui.adsabs.harvard.edu/abs/2018MNRAS.475.1093Y} {475, 1093}

\bibitem[\protect\citeauthoryear{{Zhang}, {Homeier}, {Pinfield}, {Lodieu},
  {Jones}, {Allard}  \& {Pavlenko}}{{Zhang} et~al.}{2017}]{Zhang2017}
{Zhang} Z.~H.,  {Homeier} D.,  {Pinfield} D.~J.,  {Lodieu} N.,  {Jones}
  H.~R.~A.,  {Allard} F.,   {Pavlenko} Y.~V.,  2017, \mn@doi [\mnras]
  {10.1093/mnras/stx350}, \href
  {https://ui.adsabs.harvard.edu/abs/2017MNRAS.468..261Z} {468, 261}

\makeatother
\end{thebibliography}

\end{document}